\documentclass[prb,aps,amsmath,amssymb,showpacs,floatfix,twocolumn]{revtex4-1}
\usepackage{graphicx,epsfig}
\usepackage[usenames]{color}
\newcommand{\be}{\begin{equation}}
\newcommand{\ee}{\end{equation}}
\newcommand{\bea}{\begin{eqnarray}}
\newcommand{\eea}{\end{eqnarray}}
\newcommand{\intr}{\int \mathrm{d}^3 \mathbf{r}\,}
\newcommand{\marci}{}
\newcommand{\marcitwo}{}
\newcommand{\co}{(Color online) }

\begin{document}
\title{\marci{ Global Superfluid Phase Diagram of Three Component \\
               Fermions with Magnetic Ordering} }  
\author{M. Kan\'asz-Nagy$^{1,2}$ and G. Zar\'and$^{1,2}$}
\affiliation{$^1$Freie Universit\"at Berlin, Fachbereich Physik,
  Arnimallee 14, D-14195 Berlin, Germany}
\affiliation{$^2$BME-MTA Exotic Quantum Phases 'Lend\"ulet' Group,  
Institute of Physics, Budapest University of 
Technology and Economics, Budafoki \'ut 8, H-1521 Hungary}

\begin{abstract}
We investigate a three component fermion mixture 
in the presence of  weak attractive interactions.  
We use a combination of the equation of motion and the 
Gaussian variational mean-field approaches, which  both 
allow for simultaneous superfluid and magnetic 
ordering in an unbiased way, and capture the interplay between the two 
order parameters. 
This interplay significantly modifies the phase diagram,
especially the superfluid-normal phase boundaries.
In the close vicinity of the critical temperature and for small chemical
potential imbalances,  strong particle-hole symmetry breaking leads 
to a phase diagram similar to the one predicted by Cherng {\em et al.} 
[Phys. Rev. Lett. {\bf 99},  130406 (2007)], however, the overall
phase  diagram is markedly different:  new chemical potential-driven 
first and  second order
transitions and  triple points emerge 
as well as more exotic second order
multicritical points, and 
bicritical lines with $O(2,2)$ symmetry.
We identify the terms which are necessary to capture this
complex phase diagram in  a Ginzburg-Landau approach, and  
determine the corresponding coefficients. 
\end{abstract}

\maketitle

\section{Introduction}

Experiments with ultracold atoms opened a fascinating way 
to study strong correlations and the emergence of exotic phases 
in a controlled way.\cite{Zwerger} Paradigmatic 
solid state physics models such as the fermionic and bosonic Hubbard models 
have been realized, Mott insulating and magnetic 
phases\cite{BoseMott,FermiMott} as well as \marci{various kinds of 
fermionic\cite{SfExperiment1,SfExperiment2,FFLOExp1,FFLOExp2,DensityJumpExp,Vortices}
and bosonic\cite{BoseMott,BosonicSuperfluids}}
superfluid  phases have been observed. 
{Topological excitations, e.g. vortices\cite{FFLOExp2,Vortices},
solitons\cite{Solitons}, 2D and 3D skyrmionic excitations\cite{Skyrmions}, and
knot configurations\cite{Knots} have been subjects to intensive research.
Introduction of artificial gauge fields has also been considered both 
theoretically and experimentally\cite{GaugeFields}, 
indicating that the realization of 
the quantum-Hall effect and related phenomena 
with cold atoms are within reach.}

Cold atomic systems provide, however,  not only a way to study   models
emerging in solid state physics, but they were also proposed to 
be used  to mimic  phenomena appearing in high energy and particle physics. 
In particular, attractive three component mixtures have been proposed
to simulate  quark color superfluidity\cite{HonerkampHofstetter} and 
"baryon" formation,\cite{Rapp} two fundamental concepts of 
quantum chromodynamics (QCD). An experimental realization 
of these mixtures is very difficult, but not hopeless: 
although three component systems are plagued 
by 3-particle losses,\cite{OHara,Zoller,GrimmEfimov} nevertheless, Fermi
degeneracy has been reached in  $^6$Li 
systems,\cite{ThreeCompDegenerate} which may be  just stable enough to reach  
interesting phases such as
the trionic ("baryonic") regime.\cite{OHara} Also, systems with
closed s-shells, similar to Yb  may provide an alternative and  
more stable way to realize 
almost perfectly $SU(N)$ symmetrical 
states.\cite{DemlerNatPhys,SO(N)andSU(N)Systems} 
   
In this paper, we focus on the weak coupling regime of an attractive
three component mixture, and study its low temperature color
superfluid  phases. Our main purpose is to study the effect
of chemical potential differences, and provide a complete phase 
diagram for the $SU(3)$ symmetrical interaction, which 
can be considered as the three component analogue of the famous phase
diagram of Sarma.\cite{Sarma} Surprisingly, although several studies have been
reported so far, such a phase diagram has not been 
discussed in sufficient detail so far, not even in the weak coupling
regime considered here. The first analysis 
of  Ref.~\onlinecite{HonerkampHofstetter} assumed  complete 
$SU(N)$ symmetry and has not considered the effect of different chemical
potentials. It neglected furthermore  the coupling between
ferromagnetic and superconducting order parameters. However, as 
later noticed in  Refs.~\onlinecite{Rapp} and \onlinecite{Demler}, 
$SU(3)$ symmetry allows for a coupling between 
magnetic and superfluid order parameters, and the onset of
superfluidity is therefore naturally accompanied by a ferromagnetic 
polarization\cite{Demler,InducedPolarization} 
and possibly domain formation.\cite{Rapp}  
The consequences of such coupling have been explored 
in Ref.~\onlinecite{Demler} in the immediate vicinity of the 
$SU(3)$ symmetric phase transition 
using a Ginzburg-Landau approach, however, the regime of lower
  temperatures has not been investigated.

Throughout this paper, 
we shall  proceed in the spirit of local density approximation
and  focus on a homogeneous system of three interacting fermion species,
described by the Hamiltonian 
\bea
H&=&\sum_\alpha \intr \Psi^{\dagger}_{ \alpha}(\mathbf{r}) \left(
\mathcal{H}_{0} - \mu_{ \alpha} \right) \Psi_{ \alpha}
(\mathbf{r}) 
\label{eq:H_int_def} \label{eq:H_0_def}  
\\
&-&\sum_{\alpha\ne\beta} \lambda_{\alpha\beta} \intr
\Psi^{\dagger}_{\alpha} (\mathbf{r}) \Psi^{\dagger}_{\beta} (\mathbf{r})
\Psi_{\beta} (\mathbf{r}) \Psi_{\alpha}
(\mathbf{r})\;. \nonumber
\eea
Here $\Psi^{\dagger}_{ \alpha}(\mathbf{r})$ creates a fermion 
in a hyperfine state $\alpha=1,2,3$ with  corresponding chemical potentials, 
$\mu_\alpha$. The interaction between the species is assumed 
to be local and attractive ($\lambda_{\alpha\ne\beta}>0$).\cite{footnoteInteractionStrengths}
Furthermore,  throughout most of
this work, we shall also assume $SU(3)$ symmetrical interactions, 
$\lambda_{\alpha\ne\beta}=\lambda$. This assumption is a valid  approximation 
for the $^6$Li system in the high magnetic field limit,\cite{Experimental_6Li_scattering_lengths}
and it would be certainly justified for Yb-like closed s-shell systems 
(but with attractive interactions).

This assumption is certainly
justified for Yb-like closed s-shell systems, and is  
also a valid  approximation for the $^6$Li system in the high
magnetic field limit.\cite{Experimental_6Li_scattering_lengths}  
Although the scattering lengths in the lowest 
three hyperfine states are slightly different
in the latter system, one can use 
 radio frequency and microwave fields to make them equal up to 
$\sim 0.1\%$ accuracy.\cite{Experimental_Make_sc_lengths_equal}  

The particular form of the  single
particle operator 
 $\mathcal{H}_0$ in Eq.~\eqref{eq:H_int_def} is not very important, 
since $\mathcal{H}_0$  enters the  mean-field calculations 
only through the corresponding single particle density of
states (DOS), for which we assume a simple form,  $\rho(\xi)=\rho_0
\left(1+\gamma\;\xi \right)$ and a rigid bandwidth cut-off
at $\xi=\pm W$.   Keeping the linear term $\rho_0 \gamma\,\xi$ is 
crucial: this term is the primary source of   the coupling 
between ferromagnetic and  superfluid order parameters. 
Note that in the small coupling regime  only the DOS $\rho_F$  at the 
Fermi energy and its first derivative  are expected to have 
considerable impact on the phase diagram, and therefore we do not
need to go beyond this simple linear approximation. 
We should remark though that 
the interactions renormalize the chemical potentials, and therefore 
the position of the renormalized Fermi energy, $\xi_F$ and the corresponding 
single particle density of states, $\rho_F$ must be determined 
self consistently.\cite{footnote0}

\begin{figure}[t]
\includegraphics[width=7cm,clip=true]{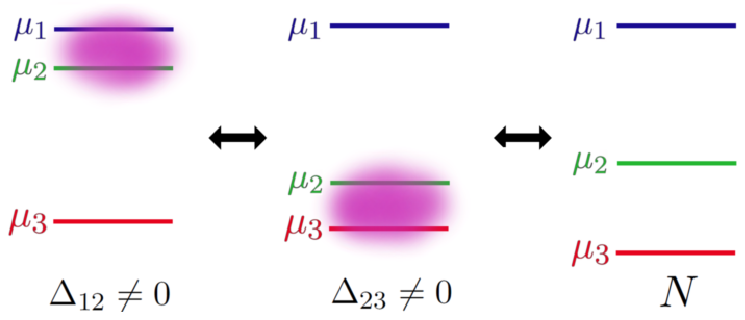}
\null
\vspace{0.8cm}
\includegraphics[width=7cm,clip=true]{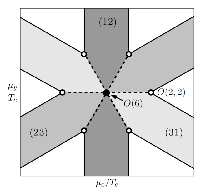}
\caption{\co Structure of the phase diagram for a constant DOS. 
Top: SF  forms between fermions with the closest chemical 
potentials, whereas at
  higher differences the normal state (N) is favored. 
Bottom: Cut of the SF phase diagram for $T \lesssim T_c$ 
Apart from the  special points (full and empty circles)
SF order always forms in one of the channels (12),
  (23) or (31). 
At the point  $\mu_x=\mu_y=0$ (full circle)
the Hamiltonian is $SU(3)$
  symmetric, and the transition is described by an $O(6)$ critical point. 
The first order lines, separating different SF phases
  terminate in second order critical points with $O(2,2)$ symmetry
  (empty circles).} 
\label{fig:Intro_schematic_phdiag} 
\end{figure}

Although we also discuss to a certain extent the role of
fluctuations in Section~\ref{sec:beyond_MF}, the bulk of this work
consists of a mean-field analysis. Even this is, 
however, not entirely trivial. In the Hubbard-Stratonovich approach of 
Refs.~\onlinecite{HonerkampHofstetterRPA} and \onlinecite{Demler} the 
decoupling of the interaction into ferromagnetic and superfluid parts
 suffers from a certain degree of arbitrariness.\cite{footnote}
Treating the ferromagnetic and superfluid order parameters at equal
footing therefore requires care.
Furthermore, at lower temperatures the second order transitions turn into 
first order transitions, and the free energy develops several
inequivalent local minima.
To cope with these difficulties, we applied two complementary 
methods:  an equation of motion method, where  vertex 
corrections are systematically neglected, and a Gaussian variational 
approach. Both approaches 
are exempt from the arbitrariness of the
Hubbard-Stratonovich transformation,  account for the interplay between
ferromagnetic and superfluid order, and, remarkably, they
both result in the {\em same} self-consistency equations. However, the Gaussian
variational approach goes beyond the equation of motion method in that 
it also  provides an estimate for the mean 
field free energy, and allows us to locate first order transitions.  
Since previous
works indicate that the 
 Fulde-Ferrell-Larkin-Ovchinnikov
(FFLO) phase with spatially varying order parameter\cite{FF,LO}
appears only in a tiny region of the phase diagram,\cite{FFLOExp1,FFLOExp2}  
here we restrict our investigation to spatially homogeneous phases.
We shall neither consider Breached Pair (BP) or Sarma
phases,\cite{Wilczek,Sarma}  since these would require fermions of 
very different masses.

Before  we turn to the more detailed presentation of the calculations,
let us summarize here our most important results.
In the small coupling limit, $T_c\ll W$, the
phase diagram is expected to become universal 
for $SU(3)$ symmetrical interactions: it should depend only 
on the dimensional  temperature, $T/T_c$, the dimensionless chemical potential
shifts, $\delta \mu_\alpha/T_c$, and the dimensionless particle-hole
symmetry breaking, $\tilde \gamma \equiv \gamma T_c$, defined in terms 
of the critical temperature $T_c$ at the $SU(3)$ symmetrical 
point, $\mu_\alpha \equiv \mu$. 
Fig.~\ref{fig:Intro_schematic_phdiag} shows the corresponding 
 schematic phase diagram in case of a particle-hole 
symmetrical situation, $\gamma=0$.
The bottom figure shows a finite temperature cut 
of the phase diagram as a function of 
the chemical potential differences, 
\bea
\mu_x&\equiv&(\mu_1-\mu_2)/\sqrt{2}\;,
\nonumber \\
\mu_y&\equiv&(\mu_1+\mu_2-2\mu_3)/\sqrt{6}\;,
\nonumber  
\eea
for a temperature $T$ fixed somewhat below the $SU(3)$ symmetrical transition
temperature, $T_c$. In the various 
gray regions two species of the smallest chemical potential difference
pair up to form a superfluid (SF) state, while the 
third species remain gapless. This explains the star-like 
structure of the phase diagram: superfluid phases appear around 
regions, where two of the chemical potentials become equal. 
 As we discuss later, 
the high ("hexagonal") symmetry of the figure
is a direct fingerprint  of the $SU(3)$ symmetrical interaction,
and a discrete particle hole symmetry. The superfluid state 
is destroyed, once all chemical potential differences become 
large compared to the condensation energy (white region). 
Close to $T_c$ the chemical potential driven SF-normal transitions are
of second order (black lines),  just as in case of a two component
mixture.\cite{Sarma} The transition between {\em different} SF  
phases is, however, always  of first order (dashed lines).

\begin{figure}[t]
\includegraphics[width=8cm,clip=true]{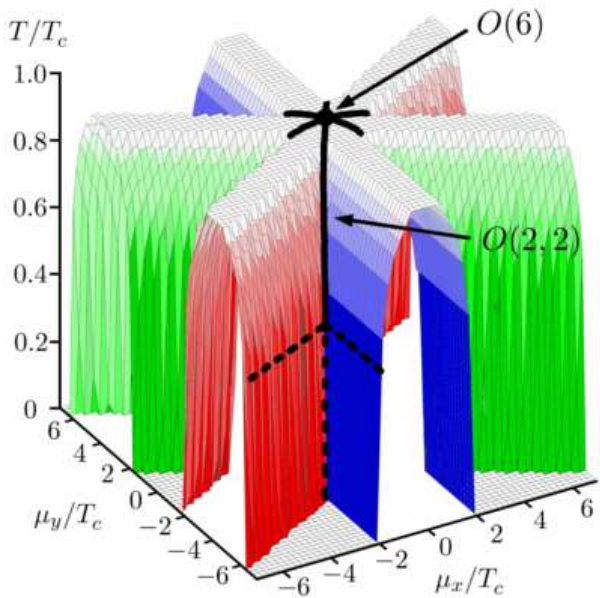}
\caption{\co  Numerically computed 
SF-N phase boundary as a function of the chemical potentials
  at constant DOS for $\lambda\rho_F= 0.1$, and $\gamma=\xi_F=0$,
  corresponding to $T_c^{SU(3)}/W=0.0076$.
The SF-N transition becomes of first order below
  a temperature $\widetilde{ T}^{\rm \;Sarma}$ (horizontal dashed lines). Vertical lines
  denote the $O(2,2)$ critical points of second (solid line) and of
  first order (dashed line). } 
\label{fig:Intro_mudiff=0.25Tc_phdiag}
\end{figure}

The phase diagram also exhibits some interesting points of special symmetry. 
At the  point $\mu_x=\mu_y=0$ the Hamiltonian is $SU(3)$ symmetrical,
and correspondingly, the phase transition at $T=T_c$ and 
 $\mu_x=\mu_y=0$ is described by
an $O(6)$ theory (the six components corresponding to the 
real and imaginary parts of the superfluid order parameters).
In three dimensions, this symmetry is spontaneously broken for $T<T_c$. 
 On the other hand, at the points indicated by white circles in 
Fig.~\ref{fig:Intro_schematic_phdiag}, the competition of two order
parameters most likely leads  
to a so-called $O(2,2)$ critical behavior (see Section~\ref{sec:beyond_MF}).

\begin{figure}[t]
\includegraphics[width=8cm,clip=true]{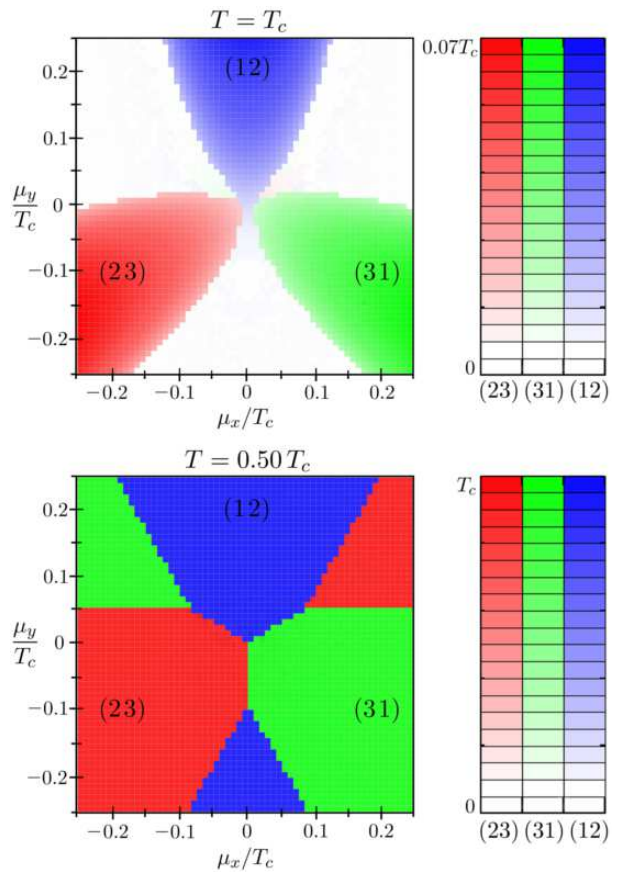}
\caption{\co Phase diagram in the vicinity of the $SU(3)$ symmetric point 
in the absence of particle-hole symmetry, $\gamma \ne 0$. 
The sixfold
  symmetry of the phase diagram is destroyed. A higher DOS can make 
SF ordering  favorable in a channel not of the smallest chemical potential
  difference, due to the gain in condensation
  energy. 
\marci{For the absolute values of the SF order parameters $\Delta_{ij}$ 
  see color code (right).} 
Parameters at the $SU(3)$ symmetric point:
  $\lambda_{\alpha\ne\beta} \rho_F=0.112, \, \gamma W=0.5, \,
  T_c/W=0.011, \xi_F/W = 0.24$ (half-filling).} 
\label{fig:Intro_mudiff=0.25Tc_alpha=0.5DOS_lambda=0.1}
\end{figure}

Fig.~\ref{fig:Intro_mudiff=0.25Tc_phdiag} shows the numerically
obtained phase diagram in a 3-dimensional plot under the assumption 
of $SU(3)$ symmetric interaction and particle-hole symmetry ($\gamma=0$). 
The dome-like structures correspond to superfluid phases with
pairing in the  $(12)$, $(23)$, and $(31)$ channels. Below the
horizontal dashed lines the chemical potential driven phase
transitions become of first order, while above these lines they are of
second order. These lines are thus the analogues of the critical point
identified by Sarma.\cite{Sarma}
 The SF-normal transitions on the 
 "roofs" of the domes belong to the $O(2)$ universality class, while 
the black solid lines correspond to $O(2,2)$ critical points. 
Finally, the crossing of the black lines at $\mu_x=\mu_y=0$
corresponds to an $O(6)$ critical point.

This rich phase diagram is further complicated if one allows for
particle-hole symmetry braking, $\gamma\ne 0$.\cite{footnote2} On
 a larger scale, 
the $\gamma\ne 0$  phase diagram looks quite similar to 
the $\gamma=0$ phase diagrams, presented in
Figs.~\ref{fig:Intro_schematic_phdiag} and 
\ref{fig:Intro_mudiff=0.25Tc_phdiag}, however, the structure of the 
phase diagram changes in the close vicinity of the 
$SU(3)$ symmetrical point. This is demonstrated in 
Fig.~\ref{fig:Intro_mudiff=0.25Tc_alpha=0.5DOS_lambda=0.1}, where the
central region of the phase diagram is shown for $T=T_c$
and $T=0.5 \,T_c$. The absence of particle-hole symmetry destroys the 
 hexagonal symmetry of the phase diagram, and leads to a trigonal 
structure, as predicted by Cherng {\em et al.}.\cite{Demler}
In this central region a higher DOS, ---  and thus 
gain in condensation   energy ---
may make SF ordering  favorable in a channel not of the smallest 
chemical potential difference. This effect is most spectacular at 
$T=T_c$, where by shifting the Fermi energy of two species one can
increase the critical temperature, and induce superfluidity
(see Fig.~\ref{fig:Intro_mudiff=0.25Tc_alpha=0.5DOS_lambda=0.1}, top).
We remark, however, that in spite of the relatively large
particle-hole asymmetry  introduced, this central region is 
typically quite small compared to the rest of the phase diagram, at
least for weak couplings, $T_c\ll W$. 
The orientation of the phases is, however, opposite to the one 
predicted in Ref.~\onlinecite{Demler}: to obtain the same 
orientation, we need to flip the sign of the slope of the DOS, and
assume a hole-like Fermi surface, $\gamma<0$. 
We must also add here that the Ginzburg-Landau action of
Ref.~\onlinecite{Demler} 
is unable to capture the endpoints of the "trigonal" region, and one
must retain higher order terms in the action to account for these
(see Section~\ref{s:Ginzburg-Landau}). 

\marci{The rest of the paper is organized as follows: In 
Section~\ref{s:Calculations}, we introduce our mean-field 
methods. We also discuss the symmetries of the order parameters,
leading to rather strong constraints on the form of the 
phase diagram. In Section~\ref{sec:MF_Phdiag}, we present
our main results on the SF phase diagram, with 
and without particle-hole symmetry, and compare our findings
with results on two component systems. In 
Section~\ref{s:Ginzburg-Landau} we present the numerical
Ginzburg-Landau expansion of the free energy around the 
$SU(3)$~symmetric point, and identify the terms responsible 
for the main features of the central part of the phase diagram.
In Section~\ref{sec:beyond_MF} we discuss the effect of 
fluctuations in the special $O(2,2)$ symmetric bicritical points. 
In Section~\ref{sec:ExpRelev} we comment on the experimental
realizability of an $SU(3)$ symmetric system. Some of the 
technical details of our calculations can be found in the 
Appendices.}

\section{Mean-field calculations} \label{s:Calculations}

In this section, we first use an imaginary time equation of motion
(EOM) method to derive the self-consistency equations for the SF and
magnetic  order
parameters. Then, to address
the low temperature regime, where these equations have
multiple solutions,\cite{Sarma} we also develop a Gaussian
variational approximation. This approach provides an estimate for 
the free energy and enables one to locate first order transitions. 

\subsection{Equation of motion technique} \label{sec:EOM_technique}

To simplify our notation, let us first 
introduce the 6 component Nambu spinor field
$$
\Phi(x)=\left( \Psi(x), \Psi^\dagger(x) \right)^T\;.
$$ 
Here we used  the compact notation
$x=(\mathbf{r},\tau)$ for the space and imaginary time
coordinates. The corresponding $6 \times 6$ propagator matrix
$\mathbf{D}(x_1,x_2) \equiv -\langle T_\tau \, \Phi(x_1) \circ
\Phi^\dagger(x_2) \rangle$ contains the normal as well as the anomalous
Green's functions of the fields $\Psi_\alpha(x)$ and,  assuming spatial
homogeneity, also obeys $\mathbf{D}(x_1,x_2)=\mathbf{D}(x_1-x_2)$. 
In order to derive
equation of motion for the propagators, we start from the imaginary
time equation of motion (EOM) of the fields, 
\be
\partial_\tau \Psi_\alpha(x)=\left[ H, \Psi_\alpha (x) \right]. \label{eq:imaginary_time}
\ee 
The EOM of the part $-\langle \mathrm{T}_\tau \Psi_\alpha (x_1)
\Psi^\dagger_\beta (x_2) \rangle$ of the propagator follows from
Eq.~\eqref{eq:imaginary_time}, and reads 
\begin{align}
\left( \partial_{\tau_1} + \mathcal{H}_0 (\mathbf{r}_1)-\mu_\alpha \right) 
  \langle \mathrm{T}_\tau \Psi_\alpha (x_1) \Psi^\dagger_\beta (x_2) \rangle  
  = \delta_{\alpha \beta} \delta_{x_1 x_2} \label{eq:EOM_propagator_example} \\ 
+ \sum_\gamma 2 \lambda_{\alpha\gamma} 
  \langle \mathrm{T}_\tau \Psi^\dagger_\gamma(x_1) \Psi_\gamma(x_1) 
  \Psi_\alpha(x_1) \Psi^\dagger_\beta(x_2) \rangle, \nonumber 
\end{align}
with $\delta_{x_1 x_2}$ denoting the four dimensional Dirac-delta
function. Similar equations hold for the anomalous propagators,
$-\langle \mathrm{T}_\tau \Psi_\alpha (x_1) \Psi_\beta (x_2)
\rangle$ and 
 $-\langle \mathrm{T}_\tau \Psi^\dagger_\alpha (x_1) \Psi^\dagger_\beta (x_2)
\rangle$.

\begin{figure}[t]
\includegraphics[width=8cm,clip=true]{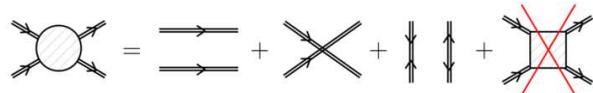}
\caption{\co Omission of the vertex corrections in the connected four
  point functions (l.h.s.). Heavy lines denote
  the full propagators, and the square stands for the vertex
  contribution.} 
\label{fig:MF_Calc_vertex_corrections}
\end{figure}

To make further progress, we simplify the four point functions
appearing in these EOMs, by simply neglecting the vertex
contribution, as shown in
Fig.~\ref{fig:MF_Calc_vertex_corrections}. This approximation is
almost equivalent to the usual BCS approximation, however, it goes
beyond that, since it allows for the simultaneous appearance of 
different kinds of order parameters in an unbiased way. 
Furthermore, even in the simple $SU(2)$ case, 
it also incorporates, e.g., the renormalization of the
Pauli susceptibility at the mean-
field level (see Section~\ref{s:2_component_SF}). 
With this approximation,
the equation of motion become solvable, and the Nambu propagator is
found to take the following form in Fourier space 
\be
\mathbf{D}(i\omega_n,\mathbf{k})^{-1}=\, i \omega_n - \mathbf{B}(\xi_\mathbf{k}). \label{eq:gapeq_Fourier}
\ee
Here $\omega_n=(2n+1)\, \pi\, T$ are fermionic Matsubara
frequencies and the matrix $\mathbf{B}(\xi)$
is defined as 
\be
\mathbf{B}(\xi) \equiv\,
  \begin{pmatrix}
    \xi - \mathbf{\Lambda} & 
    2\mathbf{\Delta} \\
    2\mathbf{\Delta}^+ & 
    -\left(\xi - \mathbf{\Lambda}^*\right)
  \end{pmatrix}. \label{eq:B_def}
\ee
The matrices
\begin{align}
\Delta_{\alpha\beta}\equiv& \,\lambda_{\alpha\beta}d_{\alpha\beta}, \label{eq:Delta_def} \\
\Lambda_{\alpha\beta}\equiv& \, (\mu_{\alpha}+2\sum_{\gamma}\lambda_{\alpha\gamma}n_{\gamma\gamma})
  \delta_{\alpha\beta} - 2\lambda_{\alpha\beta}n^*_{\alpha\beta} \label{eq:Lambda_def}
\end{align}
denote the SF order parameter\cite{footnoteOrderParametersConvDifference} and the renormalized chemical potential,
respectively. They are defined in terms of the
matrix of densities $\mathbf{n}$, and that of the  anomalous densities
$\mathbf{d}$, 
\begin{align}
n_{\alpha\beta}\equiv& \langle
\Psi_{\alpha}^\dagger(x_1)\Psi_{\beta}(x_1)\rangle, \label{eq:n_def} \\ 
d_{\alpha\beta}\equiv& \langle
\Psi_{\alpha}(x_1)\Psi_{\beta}(x_1)\rangle. \label{eq:d_def} 
\end{align}
The matrices $\mathbf{n}$ and $\mathbf{d}$ can be used to 
describe magnetic and SF
ordering, respectively. However, it is more natural to use
$\mathbf{\Lambda}$ and $\mathbf{\Delta}$ as order parameters. 
\marci{Note that, according to 
Eq.~\eqref{eq:Lambda_def}, magnetic ordering
implies a shift in the renormalized chemical potential
$\Lambda_{\alpha\beta}$, and this shift can thus also be considered 
as a magnetic order parameter. }

The expectation values Eqs.~\eqref{eq:n_def} and \eqref{eq:d_def} are
given by the propagator $\mathbf{D}(x_1-x_2)$ at equal times and equal
positions, and are thus  determined by
Eq.~\eqref{eq:gapeq_Fourier}. Taking the inverse 
of Eq.~\eqref{eq:gapeq_Fourier} and performing the Matsubara summation over
the frequencies $\omega_n$ we obtain 
\be
\int_{-W}^W \mathrm{d}\xi\, \rho(\xi)\, f \left( \mathbf{B}(\xi) \right) = 
  \begin{pmatrix}
     \mathbf{n}^*  &  -\mathbf{d} \\
     -\mathbf{d}^+  &  -\mathbf{n} + \int \mathrm{d}\xi \, \rho(\xi)
  \end{pmatrix}, \label{eq:gapeq}
\ee
where $\rho(\xi)$ denotes the DOS of $\mathcal{H}_0$, and $f$ stands
for the Fermi function.
Eqs.~(\ref{eq:B_def}-\ref{eq:Lambda_def}) 
and
\eqref{eq:gapeq} thus determine self-consistently the order
parameters $\mathbf{\Lambda}$ and $\mathbf{\Delta}$. We solve these
equations iteratively, starting from random initial conditions, and
performing the integrals in Eq.~\eqref{eq:gapeq} numerically. Notice
that $f(\mathbf{B}(\xi))$ is a matrix function, therefore, its
evaluation requires numerical diagonalization of the Hermitian matrix
$\mathbf{B}(\xi)$ for each value of $\xi$. 

We remark that the matrix $\mathbf{B}(\xi)$ in Eq.~\eqref{eq:B_def}
possesses a symplectic symmetry 
\be
\begin{pmatrix} \mathbf{0} & \mathbf{1} \\ \mathbf{1} & \mathbf{0} \end{pmatrix}
\mathbf{B}\left(\xi\right)
\begin{pmatrix} \mathbf{0} & \mathbf{1} \\ \mathbf{1} & \mathbf{0} \end{pmatrix} =
-\mathbf{B}^{\mathrm T}\left(\xi\right), \label{eq:B_symplectic}
\ee
since the order parameters $\mathbf{\Delta}$ and $\mathbf{\Lambda}$
are skew-symmetric and Hermitian, respectively. This symmetry makes the
eigenvalues of $\mathbf{B}(\xi)$ come in pairs,
$(\upsilon,-\upsilon)$, and thus simplifies some of our calculations
of the free energy in the next subsection. It is also responsible for
the structure of the equal time, equal position propagator in
Eq.~\eqref{eq:gapeq}.

\subsection{Gaussian variational approach} \label{sec:Gaussian_var}

To investigate the low temperature phase diagram, we employ a
variational method. This method consists of finding the best Gaussian
approximation to the free energy $F=-T \log \mathcal{Z}$ of the
system. As a first step, we express the grand canonical partition function
$\mathcal{Z}$ as a functional integral 
\be
\mathcal{Z} = \int \mathfrak{D} \mathbf{\overline{\psi}} \,
\mathfrak{D}\mathbf{\psi} \, 
e^ {- S[ \mathbf{\overline{\psi}}, \mathbf{\psi} ]}
, \label{eq:Z_def} 
\ee
with the action written as $S=S_0+S_\mathrm{int}$, and the
non-interacting and interacting parts defined as 
\begin{align}
S_0 &= - \frac{1}{2} \int \mathrm{d}1\,\mathrm{d}2\;
\overline{\phi}(1)\, \mathcal{D}_0^{-1}(1,2) \,
\phi(2), \label{eq:S_0_def}\\ 
S_{\mathrm{int}} &= - \sum_{\alpha\beta} \lambda_{\alpha\beta}  \int
\mathrm{d}x\; \overline{\psi}_{\alpha}(x) \overline{\psi}_{\beta}(x)
\psi_{\beta}(x) \psi_{\alpha}(x). \label{eq:S_int_def} 
\end{align}
Here $\phi=\left( \psi,\overline{\psi} \right)^T$ is a Nambu spinor
field and we used the notations $"1" = (\mathbf{r}_1,\tau_1,\nu_1)$, and $\int \mathrm{d}1 \dots$, to denote the
integration over space and imaginary time variables and the summation
over Nambu indices ($\nu_1=1, \dots, 6$) in a compact way. The inverse propagator 
\be
-\mathcal{D}_0^{-1} = \delta_{x_1 x_2}\left(\partial_{\tau_2} +
\begin{pmatrix}
\mathcal{H}_0 - \hat{\mu} & 0 \\
0 & -(\mathcal{H}_0 - \hat{\mu})
\end{pmatrix} \right), \label{eq:D_0_def} 
\ee
contains the single particle Hamiltonian of the free fields,
$\mathcal{H}_0$, where $\hat{\mu}_{\alpha\beta}=\mu_\alpha \,
\delta_{\alpha\beta}$ is a $3 \times 3$ diagonal matrix containing the chemical
potentials.

Our Gaussian approximation of the free energy is based on the standard inequality\cite{Feynman}
\be
F \leq F_G[\mathcal{D}] \equiv -T \log \mathcal{Z}_\mathcal{D} + T \langle S - S_\mathcal{D} \rangle_\mathcal{D}\;. \label{eq:Feynmans_ineq}
\ee
Here the partition function $\mathcal{Z}_\mathcal{D}$ and the average $\langle \dots \rangle_\mathcal{D}$ are defined in terms of the Gaussian action
\begin{align}
S_\mathcal{D} &\equiv - \frac{1}{2} \int \mathrm{d}1 \,\mathrm{d}2\;
\overline{\phi}(1) \, \mathcal{D}^{-1}(1,2)\,
\phi(2)\;. \label{eq:S_D_def} \\ 
\mathcal{Z}_\mathcal{D} &\equiv \int
\mathfrak{D}\mathbf{\overline{\psi}} \, \mathfrak{D}\mathbf{\psi} \,
e^{-S_\mathcal{D}[ \mathbf{\overline{\psi}}, \mathbf{\psi}
]}\;, \label{eq:Z_D_def}\\ 
\langle \dots \rangle_\mathcal{D} &\equiv \frac{1}{\mathcal{Z}_D} \int
\mathfrak{D}\mathbf{\overline{\psi}} \, \mathfrak{D}\mathbf{\psi} \;
\dots \; e^{-S_\mathcal{D}[ \mathbf{\overline{\psi}}, \mathbf{\psi}
]}\;. \label{eq:avg_def} 
\end{align}

Since we do not want to restrict our investigations to actions that
can be associated with a Hamiltonian, we do not require
$S_\mathcal{D}$ to be local. Nevertheless, at the saddle points of
$F_G$, $S_\mathcal{D}$ turns out to be local, and there exists a
Hamiltonian associated with it (see Eqs.~(\ref{eq:D=D}-\ref{eq:H_D_def}) below). 

Since the action $S_{\mathcal D}$ 
is quadratic, the propagator matrix of the Nambu fields can be written as 
\be
\mathcal{D}(1,2)=-\langle \phi(1) \, \overline{\phi}(2)
\rangle_\mathcal{D}\;, \label{eq:Grassmann_propagator} 
\ee
and expectation values can be evaluated using Wick's theorem.
We remark that the choice \eqref{eq:Grassmann_propagator} 
automatically fixes a certain ambiguity in the definition of
$\mathcal{D}^{-1}$. (For details see Appendix~\ref{App:Saddle_point}.) 
The best Gaussian approximation is given by the minimum of the
functional $F_G[\mathcal{D}]$, where $F_G$ satisfies the saddle point
equation 
\be
\frac{\delta F_G}{\delta \mathcal{D}(1,2)} = 0. \label{eq:saddle_point}
\ee
As is shown in Appendix~\ref{App:Saddle_point}, this equation is
equivalent to the self-consistency equations
(\ref{eq:B_def},\ref{eq:Delta_def},\ref{eq:Lambda_def}) and
\eqref{eq:gapeq} of the EOM technique, and amounts in  
$\mathcal{D}^{-1}$ being a local, 
\be
\mathcal{D}^{-1}(1,2)=\delta(x_1-x_2) \, \mathbf{D}^{-1}(x_2). \label{eq:D=D}
\ee
with the matrix operator on the r.h.s being just the inverse propagator
Eq.~\eqref{eq:gapeq_Fourier} in real space, 
\be
-\mathbf{D}^{-1}= \partial_{\tau_2} +
\begin{pmatrix}
\mathcal{H}_0(\mathbf{r}_2) - \mathbf{\Lambda} & 2\mathbf{\Delta} \\
 2\mathbf{\Delta}^+ & -(\mathcal{H}_0(\mathbf{r}_2) - \mathbf{\Lambda}^*)
\end{pmatrix}. \label{eq:Nambu_propagator}
\ee
The order parameters $\mathbf{\Lambda}$ and $\mathbf{\Delta}$ are
determined by the former equations,
Eqs.~(\ref{eq:Delta_def},\ref{eq:Lambda_def}). 

Thus the Gaussian
variational approach is entirely consistent with the EOM
method. However, it goes also beyond it, since it enables us to obtain
an estimate for the free energy. 
By Eqs.~\eqref{eq:D=D} and \eqref{eq:Nambu_propagator}, 
to calculate the best approximation $F_G$ to
the free energy, it is sufficient to consider local actions, for which
we can express $S_\mathcal{D}$, and thus $F_G$, in terms of a
Hamiltonian 
\be
H_\mathcal{D}=\frac{1}{2}\intr : \Phi^\dagger 
\begin{pmatrix} 
\mathcal{H}_0 - \mathbf{\Lambda} & 2\mathbf{\Delta} \\ 
2\mathbf{\Delta}^+ & -\left(\mathcal{H}_0 - \mathbf{\Lambda}^*\right)
\end{pmatrix}
 \Phi :. \label{eq:H_D_def}
\ee
Since the functional integrals are, by definition, normal ordered, the
Hamiltonian $H_\mathcal{D}$ also needs to be normal ordered, as
emphasized  by the semi-colons in Eq.~\eqref{eq:H_D_def}, indicating
normal ordering with respect to the vacuum.\cite{footnote_normal_order}

In this Hamiltonian
language, Eq.~\eqref{eq:Feynmans_ineq} takes on the form 
\be
F_G(\mathbf{\Lambda},\mathbf{\Delta})=-T \log \mathcal{Z}_\mathcal{D}
+ \langle H-H_\mathcal{D}
\rangle_\mathcal{D}\;, \label{eq:F_G_operator_formalism} 
\ee
with $H$ the full Hamiltonian of the system,
Eq.~(\ref{eq:H_0_def}),  and 
\bea
\mathcal{Z}_\mathcal{D}&=&\mathrm{Tr} e^{-\beta
  H_\mathcal{D}}, \label{Z_D_operator_formalism}
\\
\langle \dots \rangle_\mathcal{D} &=& 
{\mathrm{Tr} \left( \dots e^{-\beta H_\mathcal{D}} \right)
}/\mathcal{Z}_\mathcal{D}\;. 
\eea
Notice that $F_G(\mathbf{\Lambda},\mathbf{\Delta})$ also depends
implicitly on the chemical potentials $\mu_\alpha$ and the
temperature $T$, and it must be minimized to find the mean field value
of the variational parameters, $\mathbf{\Lambda}(\mu_\alpha,T)$ and 
$\mathbf{\Delta}(\mu_\alpha,T)$.

In this Hamiltonian approach, the evaluation of
Eq. \eqref{eq:F_G_operator_formalism} is straightforward (see Appendix
\ref{App:Calc_free_energy}), and for the free energy density we obtain 
\begin{align}
f_G=& \,\frac{1}{2} \int \mathrm{d}\xi\, \rho(\xi) \mathrm{Tr}(\xi - \mathbf{\Lambda}) \nonumber \\
-& \, \frac{T}{2} \int \mathrm{d}\xi\, \rho(\xi) \mathrm{Tr} \log
\left( 2 \cosh \left( \beta \, \mathbf{B}(\xi)/2 
\right) \right) \label{eq:f_G} \\
+& \sum_{\alpha\beta} \left(
 (\mathbf{\Lambda}_{\alpha\beta}-\mu_\alpha\delta_{\alpha\beta}
  )\mathbf{n}_{\alpha\beta}
 +\lambda_{\alpha\beta} \left(
  |\mathbf{n}_{\alpha\beta}|^2 -
  \mathbf{n}_{\alpha\alpha}\mathbf{n}_{\beta\beta} \right) 
\right) \nonumber \\ 
+& \sum_{\alpha\beta} \left(
\mathbf{\Delta}_{\alpha\beta}\mathbf{d}^*_{\alpha\beta} + 
\mathbf{\Delta}^*_{\alpha\beta}\mathbf{d}_{\alpha\beta} -
\lambda_{\alpha\beta} |\mathbf{d}_{\alpha\beta}|^2 \right) 
\nonumber.
\end{align}
Here $\beta=1/T$ is the inverse temperature, the densities
$\mathbf{n}$ and $\mathbf{d}$ are determined by Eq.~\eqref{eq:gapeq},
and the matrix $\mathbf{B}(\xi)$ is defined in Eq.~\eqref{eq:B_def}. 

As stated before, at the local minima of the functional 
$f_G$, the order parameters
$\mathbf{\Lambda}$ and $\mathbf{\Delta}$ fulfill the EOM
self-consistency equations. 
In our numerical calculations, however, we have not enforced this
constraint. Rather,  we treated the order parameters as independent
and free variables, and used  a Monte Carlo method to 
find the absolute minimum of Eq.~\eqref{eq:f_G} in the 15-dimensional
space spanned by these order parameters. In the end, we verified 
numerically that at the minima $\mathbf{\Lambda}$ and 
$\mathbf{\Delta}$ indeed satisfy the EOM self-consistency equations.

\begin{figure}[t]
\includegraphics[width=8cm,clip=true]{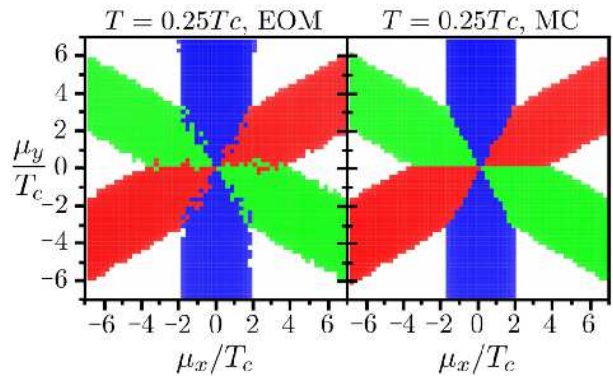}
\caption{\co Comparison of the EOM and the Gaussian variational methods. 
Left: At low temperatures the EOM equations have multiple
  solutions, and become unreliable close to first order phase
  boundaries. Right: The variational approach combined with a 
  simulated annealing identifies correctly the physically relevant
  absolute minima  of the free energy density, Eq.~\eqref{eq:f_G}. 
Parameters used were: $\lambda_{\alpha\ne\beta}
  \rho_F=0.112$, $\gamma W=0.5$, $T_c/W=0.011$, and $\xi_F/W = 0.24$
(half-filling).} 
\label{fig:Intro_T=0.25Tc_mudiff=7Tc_lambda=0.1_alpha=0.5__MCvsGP}
\end{figure}

A comparison of the variational Monte Carlo approach and the
straightforward solution of the EOM self-consistency equations is
presented in 
Fig.~\ref{fig:Intro_T=0.25Tc_mudiff=7Tc_lambda=0.1_alpha=0.5__MCvsGP}. 
At low temperatures, the EOM becomes unreliable in the vicinity of
first order phase boundaries, and finds several possible local 
minima. The variational Monte Carlo method (with simulated annealing), 
however, finds the
absolute minimum of the free energy, $f_G$, and is able to identify the
physically relevant solution.

\subsection{Symmetries} \label{sec:Symmetries}

For an $SU(3)$ symmetrical interaction,
$\lambda_{\alpha\ne\beta}=\lambda$,  the structure of the 
phase diagram is largely determined by the underlying $SU(3)$ symmetry. 
In particular, for $\mu_\alpha\equiv\mu$ the Hamiltonian is invariant
under global $SU(3)$ rotations, $\Psi_\alpha(x) \mapsto \sum_{\beta}
U_{\alpha\beta}\Psi_\beta(x)$, and a global U(1) gauge transformation, 
$\Psi_\alpha(x) \mapsto e^{i\varphi} \, \Psi_\alpha(x)$.

 The ferromagnetic order
 parameters 
$\mathbf{n}$ and $\mathbf{\Lambda}$ are Hermitian. 
They are invariant under the U(1) gauge transformation, 
and transform under $SU(3)$ rotations as 
\be
\mathbf{n}^\mathrm{T} \mapsto \mathbf{U} \, \mathbf{n}^\mathrm{T} 
\mathbf{U}^\dagger, \hspace{12 pt}
\mathbf{\Lambda} \mapsto \mathbf{U} \, \mathbf{\Lambda} \mathbf{U}^\dagger,
\label{n_trans}
\ee
which, --- after taking out the trivial  trace, --- is equivalent to the
8-dimensional adjoint representation of $SU(3)$.

The order parameters $\mathbf{d}$ and $\mathbf{\Delta}$ are, on the
other hand,  skew-symmetric, transform as 
$\mathbf{d}\mapsto e^{i2\varphi} \,\mathbf{d}$ and 
$\mathbf{\Delta}\mapsto e^{i2\varphi} \,\mathbf{\Delta}$
 under U(1) gauge transformations, and  the global $SU(3)$ group transforms
them according to
\be
\mathbf{d} \mapsto \mathbf{U} \, \mathbf{d}
\mathbf{U}^\mathrm{T}, \hspace{12 pt} 
\mathbf{\Delta} \mapsto \mathbf{U} \, \mathbf{\Delta} \mathbf{U}^\mathrm{T}, 
\label{d_trans}
\ee
which is equivalent to the conjugate representation of $SU(3)$. This
can be seen by introducing the 3 component vectors
$\underline{d}_\alpha = \frac{1}{2} \sum_{\beta\gamma}
\epsilon_{\alpha\beta\gamma} d_{\beta\gamma}$ and 
$\underline{\Delta} = \frac{1}{2} \sum_{\beta\gamma} \epsilon_{\alpha\beta\gamma}
\Delta_{\beta\gamma}$ by means of the completely antisymmetric
Levi-Civita symbol $ \epsilon_{\alpha\beta\gamma}$. In this form  
Eq.~\eqref{d_trans}
reads 
\be
\underline{\mathbf{d}} \mapsto \, \mathbf{U}^*
\underline{\mathbf{d}}, \phantom{nn} 
\underline{\mathbf{\Delta}} \mapsto  \mathbf{U}^*
\underline{\mathbf{\Delta}}\;.  
\ee

In the special case, $\lambda_{\alpha\ne\beta}=\lambda$ and
$\mu_\alpha=\mu$, symmetry implies that the Ginzburg-Landau
functional must be invariant under the transformations 
\eqref{n_trans} and \eqref{d_trans}, and the U(1) gauge 
transformation. The onset of  superfluidity, however, spontaneously
breaks the $SU(3)\otimes U(1)$ symmetry down  to $SU(2)\otimes U(1)$.
This spontaneous symmetry breaking is accompanied by 
the emergence of five Goldstone modes.\cite{HonerkampHofstetterRPA}

The presence of the chemical potentials, 
$\hat \mu_{\alpha\beta} = \delta_{\alpha\beta}\mu_\alpha$, obviously breaks the $SU(3)$
symmetry. However, one has strong symmetry-dictated 
constraints on the Ginzburg-Landau functional even in this case, and
the latter must be invariant with respect to the transformations
in Eqs.~\eqref{n_trans} and
\eqref{d_trans},  provided
that $\hat \mu $ is also transformed accordingly,  
$\mathbf{\hat \mu} \mapsto \mathbf{U} \, \mathbf{\hat \mu}
\mathbf{U}^\dagger$ (see also Section~\ref{s:Ginzburg-Landau}). 
In addition, even in the presence of chemical potential differences, 
$SU(3)$ symmetry implies Ward identities,\cite{Demler} relating four-point
expectation values and the ferromagnetic order parameter
$\mathbf{n}$ as 
\be
\left( \mu_\alpha - \mu_\beta \right) n_{\alpha\beta} = \sum_\gamma	
  2 \left( \lambda_{\beta\gamma} - \lambda_{\alpha\gamma} \right) 
  \langle {\Psi}^\dagger_\gamma {\Psi}^\dagger_\alpha 
  \Psi_\beta \Psi_\gamma \rangle\;. 
\label{Ward}
\ee 
From this identity (derived in Appendix~\ref{sec:Ward1}) 
it follows that $\mathbf{n}$ is  diagonal for an $SU(3)$ 
symmetric interaction. We remark that a similar approximate 
Ward identity can be derived 
within the Gaussian variational method (see Appendix~\ref{sec:Ward2}),
leading to the same conclusions.
 
The off-diagonal elements of the chemical potential tensor, 
$\mathbf{\hat \mu}$ describe tunneling between different hyperfine
components, and they typically vanish  in practical situations. 
Under these restrictions, allowed $SU(3)$ rotations generate 
essentially only permutations of the hyperfine labels, 
$\alpha$, and the corresponding chemical potentials, $\mu_\alpha$. 
On the $(\mu_x,\mu_y)$ plane,  these permutations translate 
to $C_3$ rotations and reflections, and give a two-dimensional  
representation of the ${\mathbb S}_3\sim C_{3v}$ group, implying a
{\em triangular} symmetry of the phase diagram in this plane 
(see Fig~\ref{fig:Intro_mudiff=0.25Tc_alpha=0.5DOS_lambda=0.1}).

In addition to the symmetries discussed above, 
for an $SU(3)$ symmetrical Hamiltonian,  the mean field 
equations also have a certain {\em particle-hole symmetry}
if the single particle density of states obeys 
$\varrho(\xi) = \varrho(-\xi)$, and the chemical potentials are set 
to a value, $\mu\to\mu_{\rm half}$,
 such that $\varrho$ is exactly half-filled. 
Under these 
conditions we can show (see Appendix~\ref{appendix_ph}) that the mean
field solutions are  symmetrical in the sense that for $\delta\mu_\alpha\equiv
\mu_\alpha-\mu_{\rm half}$ and for $\delta\mu_\alpha\to -\delta\mu_\alpha$ the superfluid 
and magnetic symmetries are broken in the same 
channels  and the order parameters are also 
equal apart from signs, global gauge transformations, and conjugation.
In this special case, due to the additional permutational 
symmetry discussed above,
the phase diagram exhibits a {\em sixfold $C_{6v}$ symmetry} in the 
$(\delta\mu_x,\delta\mu_y)$ plane for traceless chemical potential shifts, 
$\delta\mu_1 +\delta\mu_2 + \delta\mu_3 =0$, (see Fig.~\ref{fig:Intro_schematic_phdiag}.) 

This particle-hole symmetry also emerges at the  level of the Hamiltonian 
in certain cases.  The half-filled attractive three component 
Hubbard model on a bipartite  lattice  
$$
H = -t \sum_\alpha \sum_{\langle ij\rangle }
(a^\dagger_{i\alpha} a_{j\alpha} + h.c.) - \frac U 2 \sum_i 
(\sum_\alpha n_\alpha -\frac3 2)^2\;,
$$
e.g., has an exact particle-hole symmetry: it is invariant under 
the unitary transformation $a_{i\alpha} \leftrightarrow
\mathrm{sign}(i)\;a^\dagger_{i\alpha}$, with $\mathrm{sign}(i)$
taking values $\pm $ for the two sublattices. Just as the mean field
symmetry discussed in the previous paragraph, 
this exact symmetry 
relates the order parameters of the 
symmetry broken phases for $\pm\delta\mu_\alpha$.
We remark that, on a lattice, for stronger couplings, 
in addition to the SF/magnetic phases discussed here,
other non-trivial phases may emerge (eg. charge density waves
 or trionic phases).\cite{Rapp,InducedPolarization}

Although the particle-hole  symmetry discussed here holds only for a
single and  special chemical potential
value, we found that for $T_c\ll W$ higher order terms in the 
 Ginzburg-Landau action are only sensitive to the immediate 
vicinity of the Fermi surface. As a result, particle-hole symmetry 
becomes an {\em approximate symmetry} with a good accuracy, 
whenever the slope of the 
single particle density of states vanishes, $\gamma \equiv 0$. 
For $\mu_\alpha\equiv \mu$, 
$\lambda_{\alpha\ne \beta}\equiv \lambda$, and 
$\gamma \equiv 0$ we thus recover a phase diagram of 
hexagonal symmetry within our numerical accuracy 
(see Fig.~\ref{fig:Intro_schematic_phdiag}).

\section{Mean-field phase diagram} \label{sec:MF_Phdiag}

Let us now present the phase diagrams in the weak coupling limit,
$T_c \ll W$, as obtained
numerically, by the EOM and Monte Carlo methods presented in Section
\ref{s:Calculations}.

\subsection{Constant density of states ($\gamma=0$)} \label{sec:Constant_DOS}
As we argued in the Introduction, except for the $SU(3)$ symmetric
point, a system of constant DOS always favors the formation of a SF phase in one of
the pairing channels (12), (23) and (31), having 
the smallest chemical potential difference. If the chemical
potential difference between the components forming the SF state
exceeds a certain limit (known as the Clogston limit\cite{Clogston_limit}
at zero temperature in case of two fermionic components),
the system goes into the normal phase. This transition can either be
of first or of second order, depending on the temperature.\cite{Sarma}

\begin{figure}[t]
\includegraphics[width=8cm,clip=true]{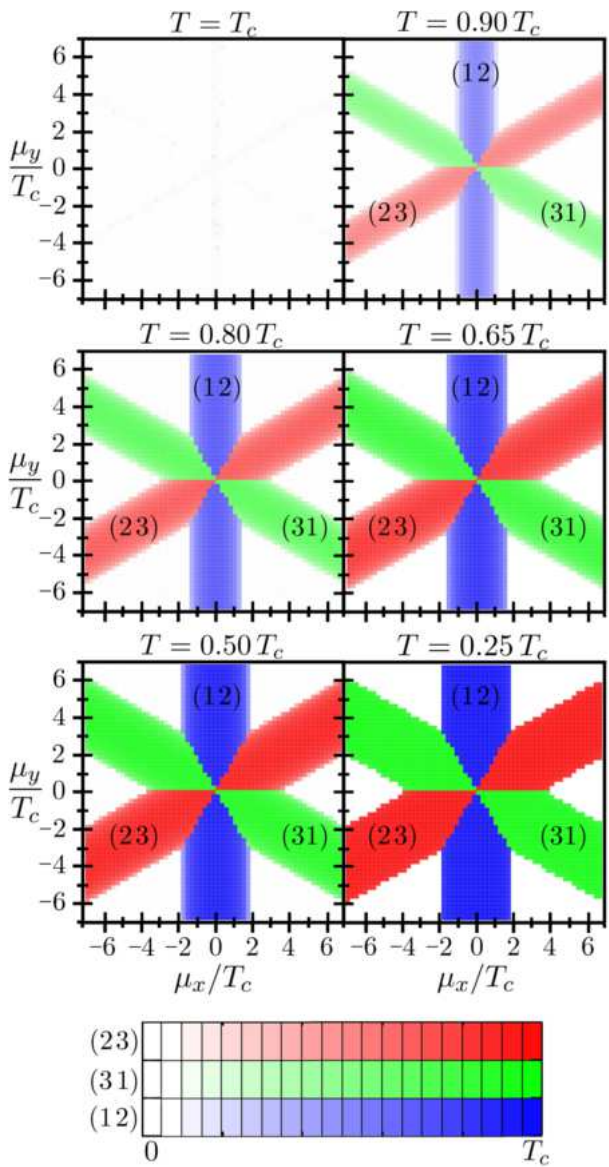}  
\caption{\co  Mean-field phase diagrams at constant DOS, $\gamma=0$. 
Different SF   phases are separated by first order lines. 
At $T=0.25\,T_c$ SF-N transitions are of first
  order, whereas they become of second order 
  for $T>\widetilde{T}^{\rm Sarma} \approx 0.48 T_c$ 
  (see Section~\ref{s:2_component_SF}).
Absolute values of components of the order parameter 
  $\mathbf{\Delta}$ are given 
  in units of $T_c$ (see color code). 
Parameters at the $SU(3)$ symmetric point:
  $\lambda_{\alpha\ne\beta} \rho_F =0.1$, $T_c/W=0.0076$, 
  and $\gamma W=\xi_F/W=0$.} 
\label{fig:MF_Phdiag_mudiff=7Tc_alpha=0DOS_lambda=0.1}
\end{figure}

Fig.~\ref{fig:MF_Phdiag_mudiff=7Tc_alpha=0DOS_lambda=0.1} shows the
numerically obtained phase diagram at different temperatures. 
All these cuts have the structure presented in 
Fig.~\ref{fig:Intro_schematic_phdiag}. The hexagonal symmetry of the
middle of the phase diagram is related to $SU(3)$ symmetry:
it is due to the invariance of the
Hamiltonian under the permutations of the fermion species ($\alpha
\leftrightarrow \beta$ and 
$\mu_\alpha \leftrightarrow \mu_\beta$)  
and the approximate particle-hole symmetry, as explained in 
Section~\ref{sec:Symmetries}.
The first order SF-SF transitions
appear along lines where the chemical potential differences
between two different pairs of fermions become equal. 
Along some special
directions 
in  the $(\mu_x, \mu_y)$ plane 
two out of three fermions have equal
chemical potentials, and can form a SF state
even far away from the central $SU(3)$ symmetric
point. This explains the ray-like structures in
Fig.~\ref{fig:MF_Phdiag_mudiff=7Tc_alpha=0DOS_lambda=0.1}. In all 
other directions the chemical potential differences
continue to grow until the system goes into the normal 
phase at chemical potential differences of the order of 
the superfluid gap at the $SU(3)$ symmetric point. 
For $T> \widetilde{T}^{\rm Sarma}\approx 0.48 \, T_c$ 
this  chemical potential driven SF-normal transition
is of second order, however it becomes of first order below
$ \widetilde{T}^{\rm Sarma}$ (see Section~\ref{s:2_component_SF}).

\subsection{Linear density of states ($\gamma \ne 0$)} \label{sec:Linear_DOS}

\begin{figure}[t]
\includegraphics[width=8cm,clip=true]{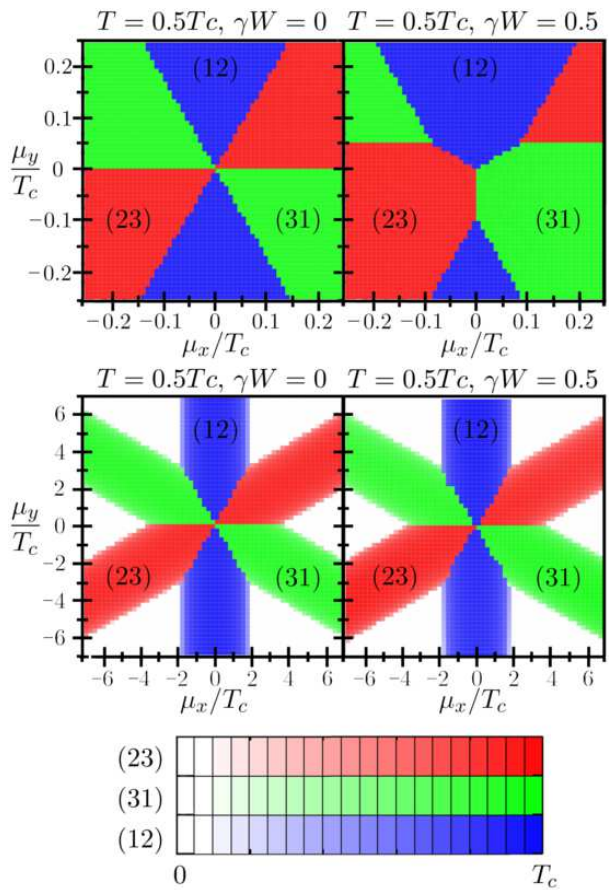}
\caption{\co  Phase diagrams at constant (left) and linear (right) DOS at
  $T=0.5 T_c$. Non-zero $\gamma$ deforms the middle of the phase
  diagram (top right), whereas on the large scale (middle), the phase
  diagram with constant and linear DOS are almost
  identical. Parameters at the $SU(3)$ symmetric point:
  $(\lambda_{\alpha\ne\beta} \rho_F, \, T_c/W, \, \gamma W, \xi_F/W) =
  (0.1, 0.0076, 0, 0)$ in the left and $(0.112,0.011, 0.5, 0.24)$ in
  the right figures.  } 
\label{fig:MF_Phdiag_T=0.5Tc_alpha=0_0.5}
\end{figure}

In case of a non-constant DOS ($\gamma \ne 0$), particle-hole symmetry
is broken at the Fermi surface, even at the $SU(3)$ symmetric point. 
At a first glance, the phase diagram is only slightly different from
the $\gamma = 0$ case,
however, at a closer look qualitative differences can be
discovered (see bottom and top parts of 
Fig.~\ref{fig:MF_Phdiag_T=0.5Tc_alpha=0_0.5}). 
For $\gamma \ne 0$ the SF state not necessarily forms in
the channel with the smallest chemical potential difference. The
reason is that the gap is exponentially sensitive to the DOS. 
As a result, it may be favorable to form an SF state 
in channels, where the DOS is larger at the 
chemical potential, even at the expense of Zeeman energy 
(chemical potential) loss.
 This mechanism is driven by the derivative
of the DOS $\gamma$, and changes the phase diagram close to the 
$SU(3)$ symmetric point. Here the phase diagram has
only three-fold symmetry, corresponding to 'color' permutations, and
superfluidity forms in channels of the largest density of states. At
higher values of the chemical potential, however, the phase diagram 
remains essentially  unaltered, and is 
almost identical to that of constant density of states.

These results are similar to the predictions of 
Ref.~\onlinecite{Demler}, however, the phase structure 
differs somewhat, and the direction of 
the phase diagram of Ref.~\onlinecite{Demler}
seems to be flipped. We verified,
that both  the variational calculation and the equation of
motion method yield consistently the phase diagram presented
here,  
which we can also reproduce by the Ginzburg-Landau approach, 
presented in Section~\ref{s:Ginzburg-Landau}. As we discuss 
there, the Ginzburg-Landau action of 
Ref.~\onlinecite{Demler} cannot produce the six-fold symmetric structure of
the overall phase diagram, and one needs to keep higher order terms
to recover it.

The previously discussed region of three-fold symmetry is, however,
usually small compared to the overall scale of the phase diagram.  
For the parameters \marci{of the left figures in}
Fig.~\ref{fig:MF_Phdiag_T=0.5Tc_alpha=0_0.5}, e.g., 
 $T_c/W=0.011$, and a relatively steep density of 
states with $\gamma W = 0.5$, the
three-fold symmetric region is present only
for $|\mu_x|,|\mu_y| <  0.1\;T_c$, while the overall scale of the 
phase diagram is about $\sim 3\;T_c$. The relative 
size of this central region increases for larger interaction strengths, and  
for $T_c/W=0.105$ and $\gamma W = 0.5$ we find, e.g.,
that the central triangular region
extends to $|\mu_x|,|\mu_y| <  0.25\;T_c$. The size of the central
triangular region seems to scale roughly as $\sim \sqrt{\gamma T_c}$.

\begin{figure}[h]
\includegraphics[width=8cm,clip=true]{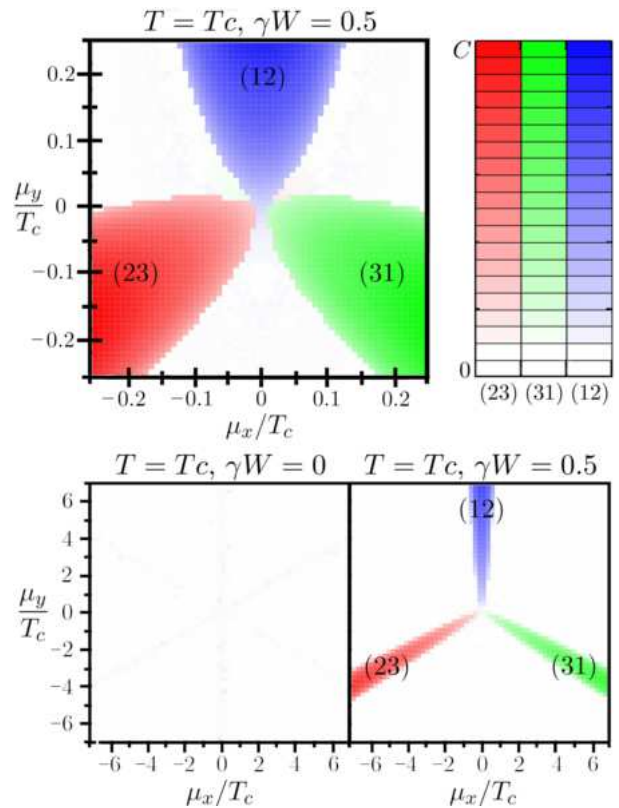}
\caption{\co  Phase diagrams at $T=T_c$ with constant (bottom left) and
  linear ($\gamma W = 0.5$) DOS (top left and bottom right). For
  linear DOS the SF-N critical temperature can exceed $T_c$ of the $SU(3)$
  symmetric point, 
whereas for $\gamma=0$, the SF
  phase disappears everywhere above $T_c$. 
The largest values of the color scales correspond to 
 $|\Delta_{\alpha\beta}| =0.07 T_c$ (top left), 
and $|\Delta_{\alpha\beta}|=T_c$ (bottom left and
  right). Parameters at the $SU(3)$ symmetric point are
  $(\lambda_{\alpha\ne\beta} \rho_F, \, T_c/W, \, \gamma W, \xi_F/W) =
  (0.1, 0.0076, 0, 0)$ in the bottom left and $(0.112, 0.011, 0.5,
  0.24)$ in the top left and bottom right figures. } 
\label{fig:MF_Phdiag_T=Tc_alpha=0.5}
\end{figure}

In Fig.~\ref{fig:MF_Phdiag_T=Tc_alpha=0.5} we confirm the predictions
of Ref.~\onlinecite{Demler}, that breaking the $SU(3)$ symmetry by
the chemical potential can indeed lead to the appearance of
superfluidity. Again, this is simply related to the fact, that 
the superfluid transition temperature is exponentially sensitive to 
the DOS at the Fermi energy.  At the $SU(3)$ critical temperature $T_c$,
superfluidity appears only in small regions of the phase diagram,
around the lines where two of the three fermion species have equal
chemical potentials. These regions lie on that side of the $SU(3)$
symmetric point, where the particles of the closest chemical
potentials have higher DOS at the Fermi energy than the third
one. We remark that the expansion of the free energy up
to third order in the order parameters can not recover this structure
precisely, and here the phase diagram is  significantly different 
from the phase diagram of Ref.~\onlinecite{Demler}.

\subsection{Two component superfluidity} 
\label{s:2_component_SF}

It is instructive to compare our mean-field theory with results 
obtained for two component systems. 
As first noticed by Sarma,\cite{Sarma}  for two component systems
the Zeeman field-induced SF-N transition becomes of 
first order below  the temperature
${T}^{\rm Sarma}$, and above the chemical potential difference
${\mu}_x^{\rm Sarma}=({\mu}_1^{\rm Sarma}-{\mu}_2^{\rm
  Sarma})/\sqrt{2}$. 
Sarma also determined the mean-field values of this critical point
(Sarma point), and obtained
\be
{T}^{\rm Sarma} =  \, 0.58 \,
{T}_c\;, \phantom{nnn}
{\mu}_x^{\rm Sarma} =  \, 1.5 \, {T}_c\;, 
\label{eq:Sarma_point_Sarma} 
\ee
with ${T}_c$ the critical temperature at $\mu_x=0$. 
He also determined the critical chemical potential difference
at zero temperature, known as the Clogston limit\cite{Clogston_limit}, 
\begin{equation} 
{\mu}^{\rm Clog}_x = 2 \, {\Delta}(T=0)= 1.764 \, T_c\;, 
\label{eq:Clogston_limit_Sarma}
\end{equation}
with ${\Delta}$ denoting  the SF order parameter.

\begin{figure}[t]
\includegraphics[width=8cm,clip=true]{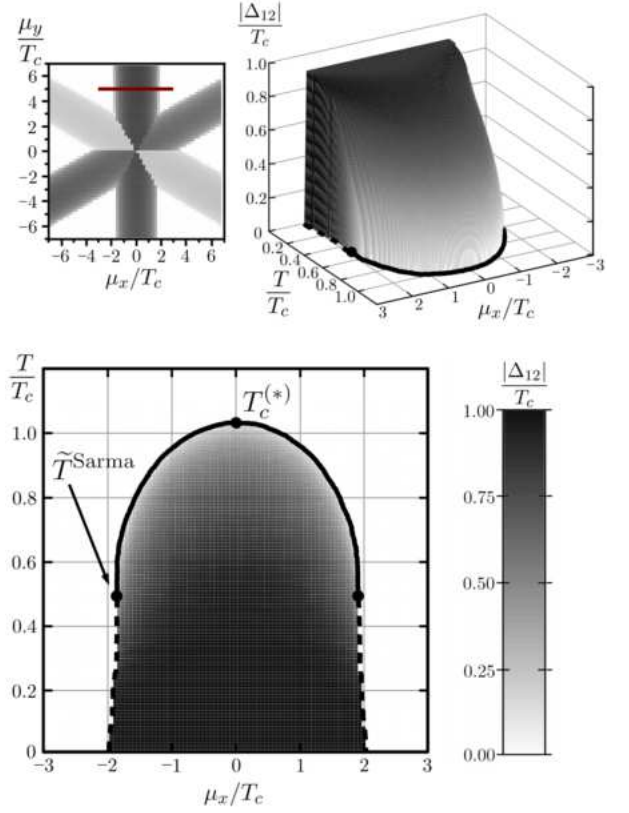}
\caption{\co SF phase diagram (top right and bottom) at linear DOS
  ($\gamma W = 0.5$), with $\mu_y=5 T_c$ kept constant, as indicated
  by the solid line in the top left figure. The SF-N transition
  becomes from second order (solid line) to
 first order (dashed line) below  the temperature 
$\widetilde{T}^{\rm Sarma}=0.48 \, T_c^{(*)}$, 
and chemical potential difference $\widetilde{\mu_x}^{\rm Sarma} 
= 1.842 \, T_c^{(*)}$, with $T_c^{(*)}=1.027 \, T_c$ 
the critical temperature at $\mu_x=0$ and $\mu_y=5 T_c$. Parameters at
the $SU(3)$ symmetric point were: $ \lambda_{\alpha\ne\beta}\rho_F=0.1,
\gamma W = 0.5, T_c/W = 0.0076, \xi_F=0$.} 
\label{fig:MF_Phdiag_T_mux_lambda=0.1_alpha=0.5_muy=5Tc_mudiffx=3Tc_T0_1.1TcSU3__absD12} 
\end{figure}

The three component system 
exhibits a two component behavior in regimes where the
chemical potential of two species remains close,
e.g. $|\mu_1-\mu_2| \sim T_c$, while that of the third component is
very far from them ($|\mu_3|\gg T_c$).
 To investigate this limit, we fixed $\mu_y=5 T_c$, and
varied  $\mu_x$, along the solid line shown in
the top left panel of
Fig.~\ref{fig:MF_Phdiag_T_mux_lambda=0.1_alpha=0.5_muy=5Tc_mudiffx=3Tc_T0_1.1TcSU3__absD12}.
The corresponding SF phase diagram displays features similar 
to those predicted by Sarma. 
At $T=0$ temperature, the absolute value of the SF order
parameter is independent  of $\mu_x$ in the superfluid phase, 
and its magnitude agrees
with the BCS result, $\Delta(T=0) =0.882 \; T_c^{(*)}$,
with $T_c^{(*)}$ being  the critical temperature at $\mu_x=0$.\cite{footnoteTcStar} 
The critical value of $\mu_x$ (Clogston limit),
however, shows significant deviations compared to 
Eq.~\eqref{eq:Clogston_limit_Sarma}. 
For a coupling  $\tilde{\lambda} \equiv \lambda \rho_F = 0.1$, e.g., 
we find both for a two \textit{and} for a three component
system 
\begin{equation}
\mu^{\rm Clog}_{x} \to  \left. \widetilde{\mu}^{\rm Clog}_{x} 
\right|_{\tilde \lambda = 0.1}= 
2.19 \, \Delta(T=0) = 1.93 \, T_c^{(*)}\;. \label{eq:Clogston_limit_KNMZG1}
\end{equation}
For $T_c\ll W$, the prefactor was found to be approximately independent
 of the value of $\mu_y$ and particle-hole
  symmetry breaking parameter, $\tilde \gamma$.
The difference between Eq.~\eqref{eq:Clogston_limit_KNMZG1} 
and Clogston's result is due to the inclusion of magnetic degrees of freedom
in the free energy density, Eq.~\eqref{eq:f_G}, 
which accounts for interaction-related 
contributions to the  Pauli susceptibility, 
$\chi\sim \rho_F$, neglected in Sarma's
work.\cite{Sarma} These susceptibility contributions are proportional to
$\lambda_{\alpha \beta}\rho_F^2$, and therefore result
in a correction to the magnetic energy 
of relative size  $\sim \lambda_{12}\rho_F$, in rough agreement  
with the numerically observed shift of $\mu^{\rm Clog}_{x}$.
It is easy to understand this difference 
on physical grounds: In the SF state (12), 
the densities $n_{11}$ and $n_{22}$ are exactly equal 
at $T=0$,  while  in the  normal  state they 
shift according to the chemical potential
difference. The interaction is, however, repulsive in the 
magnetic channel. Consequently, the (magnetized) normal state 
becomes less favorable, and $\mu^{\rm Clog}_{x}$ shifts upwards.

Locating numerically the Sarma point we also find that 
it is shifted compared to Eq.~\eqref {eq:Sarma_point_Sarma}, 
\begin{eqnarray}
{T}^{\rm Sarma}\to \left. \widetilde{T}^{\rm\; Sarma} \right|_{ \tilde{\lambda} = 0.1}
  = & \, 0.48 \, T_c^{(*)}\;, \label{eq:Sarma_point_KNMZG1} \\
{\mu}_x^{\rm Sarma}\to \left. \widetilde{\mu}_x^{\rm \;Sarma} \right|_{ \tilde{\lambda} = 0.1} 
  = & \, 1.842 \, T_c^{(*)}\;, \label{eq:Sarma_point_KNMZG2}
\end{eqnarray}
again, approximately independently from the value of $\gamma$.
These results and 
Eq. \eqref{eq:Clogston_limit_KNMZG1}
demonstrate that the  positions of the Sarma point and 
the Clogston point, Eq.~\eqref{eq:Clogston_limit_Sarma} can significantly 
deviate from their standard BCS values due to interaction effects.
Furthermore, their independence from the particular value of $\gamma$
shows that, at least for $T_c\ll W$, particle-hole symmetry breaking 
does not have a significant effect on the SF phases in the regime 
where the chemical potentials are
far from the $SU(3)$ symmetric point.

\begin{figure}[h]
\includegraphics[width=7.5cm,clip=true]{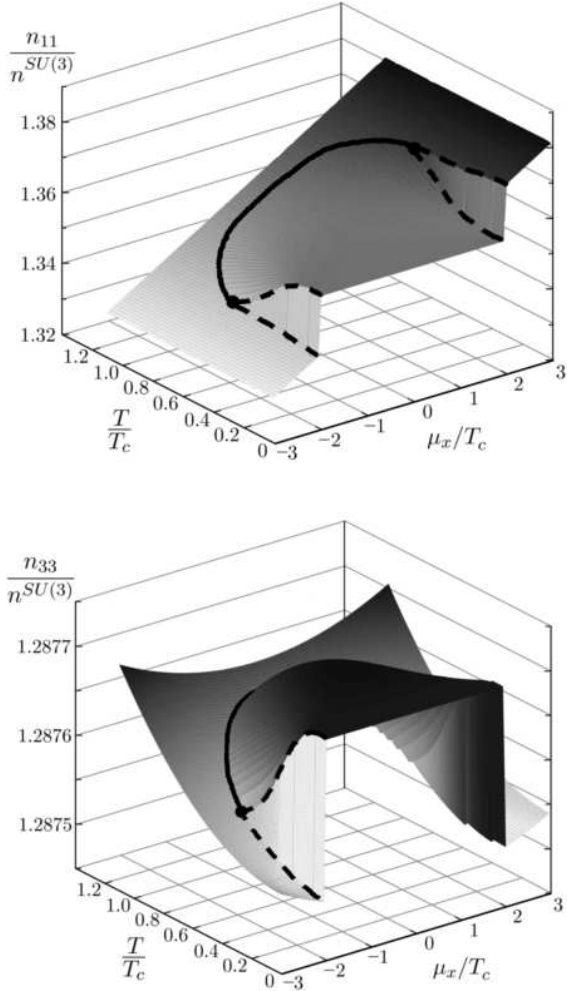}
\caption{\co Interplay between superfluidity and magnetism in the SF
  channels, $\alpha=1,2$ (top), and for the third, normal component (bottom), for linear DOS
  ($\gamma W = 0.5$), with $\mu_y=5 T_c$ kept constant. The shift of
  the densities along the SF-N phase boundary is smooth for
  $T>\widetilde{ T}^{\rm \;Sarma}$ (solid line), and discontinuous for 
  $T<\widetilde{ T}^{\rm \;Sarma}$ (dashed  line). 
  The density jump of the third component is much smaller than
  that of the SF components.
 [See also
    Fig.~\ref{fig:MF_Phdiag_T_mux_lambda=0.1_alpha=0.5_muy=5Tc_mudiffx=3Tc_T0_1.1TcSU3__absD12}.]
  Parameters at the $SU(3)$ symmetric point: $
  \lambda_{\alpha\ne\beta}\rho_F=0.1, \gamma W = 0.5, T_c/W =
  0.0076, \xi_F=0$.} 
\label{fig:MF_Phdiag_T_mux_lambda=0.1_alpha=0.5_muy=5Tc_mudiffx=3Tc_T0_1.1TcSU3__n}
\end{figure}

In the SF state,  the SF species are bound together, and the 
condensate itself cannot be polarized. This has
an experimentally important manifestation at the SF-N transition,
where a sudden shift appears in 
the densities at the phase boundary, as presented in
Fig.~\ref{fig:MF_Phdiag_T_mux_lambda=0.1_alpha=0.5_muy=5Tc_mudiffx=3Tc_T0_1.1TcSU3__n}. At
zero temperature, the densities in the SF channel are equal, and their
value does not depend on the chemical potential difference, whereas at
the SF-N transition, a difference in the densities sets in. At
temperatures below $\widetilde{ T}^{\rm \;Sarma}$, the SF-N transition is of first
order, and the densities jump discontinuously on the phase boundary. In
Fig.~\ref{fig:MF_Phdiag_T_mux_lambda=0.1_alpha=0.5_muy=5Tc_mudiffx=3Tc_T0_1.1TcSU3__n} this
amounts to a $\sim 1 \%$ jump in the densities. 
In the strongly interacting
regime, however, the jump is expected to take 
much higher values, similar to two component
systems.\cite{DensityJumpExp} 

Let us close this section by investigating
 the effect of SF transition on the third, normal component. 
Indeed, in the presence particle-hole symmetry breaking, 
the SF order parameter couples directly to the magnetization, and
should shift the density of the third component.
Fig.~\ref{fig:MF_Phdiag_T_mux_lambda=0.1_alpha=0.5_muy=5Tc_mudiffx=3Tc_T0_1.1TcSU3__n}
shows this effect 
for a linear DOS in the weak coupling limit. 
We find that the shift in the density of  the third component is only 
 of the order  of $0.01 \%$ for  $T_c/W=0.0076$, however, for larger
 ratios, $T_c/W = 0.1$ (but the same $\gamma$) it reaches values of 
 the order of $1 \%$, indicating that this effect
may be measurable in the strong coupling regime.

\section{Ginzburg-Landau action} \label{s:Ginzburg-Landau}

In this section, we focus on the central region of the phase diagram,
and construct  a Ginzburg-Landau (GL) expansion of the free energy
\eqref{eq:f_G} around the $SU(3)$-symmetric point, $\mu_x=\mu_y=0$ for
 $T\approx T_c$. Throughout this section, 
we assume a perfectly $SU(3)$ symmetrical interaction, 
$\lambda_{\alpha\ne\beta}=\lambda$.  While the form of the Ginzburg-Landau
functional is dictated by symmetry, the coefficients of the various
terms depend on the microscopic parameters. We shall give approximate 
expressions for them, as obtained through a numerical 
analysis of Eq.~\eqref{eq:f_G}.
 
In the weak coupling limit, the dimensionless free energy density,
$$
\widetilde{f}_G\equiv f_{G}/(\rho_F T_c^2)\;,
$$
can only depend on a 
few dimensionless physical parameters:  the dimensionless interaction 
$\widetilde{\lambda}\equiv\rho_F \lambda$,
the dimensionless slope of the DOS at the Fermi energy $\widetilde{\gamma} \equiv
\gamma T_c$,  
the reduced temperature $t \equiv (T-T_c)/T_c$, and 
 the dimensionless chemical
potential differences $\delta\widetilde{\mu}=(\mu-\mu^{SU(3)})/T_c$, 
with  $\mu^{SU(3)}$ denoting the  chemical potential at the
$SU(3)$ symmetric point.\cite{footnote3}  
Most importantly, however, $\widetilde{f}_G$ is a functional of the
dimensionless order  parameters, 
\bea 
\widetilde{\mathbf{\Delta}}\equiv\mathbf{\Delta}/T_c\;,\phantom{nn}
\delta\widetilde{\mathbf{\Lambda}}\equiv(\mathbf{\Lambda}-\mathbf{\Lambda}^{SU(3)})/T_c\,, 
\eea
with $\mathbf{\Lambda}^{SU(3)}$ denoting the  renormalized chemical potential
at the $SU(3)$ symmetric point.

The expansion of the free energy contains only $SU(3)$ invariant terms
and can therefore be expanded as\cite{Demler}
\begin{align}
\widetilde{f}_G &=
A_1\; \mathrm{Tr} (
\widetilde{\mathbf{\Delta}}\widetilde{\mathbf{\Delta}}^+ ) + A_2\, \mathrm{Tr}
(
(\widetilde{\mathbf{\Delta}}\widetilde{\mathbf{\Delta}}^+)^2) 
\label{eq:GL_SU(3)_third_order}
\\  
&+
  \,B_1\,\mathrm{Tr} (
\delta\widetilde{\mathbf{\Lambda}}^2 ) + B_2\, \mathrm{Tr} (
\delta\widetilde{\mathbf{\Lambda}} )^2 + B_3\, \mathrm{Tr} (
\delta\widetilde{\mathbf{\mu}}\;\delta\widetilde{\mathbf{\Lambda}} ) \nonumber
\\ 
&+C_1\, \mathrm{Tr} (
\delta\widetilde{\mathbf{\Lambda}}\widetilde{\mathbf{\Delta}}\widetilde{\mathbf{\Delta}}^+) 
+ C_2\, \mathrm{Tr} ( \delta\widetilde{\mathbf{\Lambda}} ) \mathrm{Tr} ( 
\widetilde{\mathbf{\Delta}}\widetilde{\mathbf{\Delta}}^+ ) \nonumber \\ 
&+ C_3\, \mathrm{Tr} (
\delta\widetilde{\mathbf{\mu}}\widetilde{\mathbf{\Delta}}\widetilde{\mathbf{\Delta}}^+)
+ \dots . \nonumber 
\end{align}
The 8 coefficients appearing in this expansion 
are all functions of $\widetilde{\lambda}$, $t$, and
$\widetilde{\gamma}$. We determined them by fitting
the free energy Eq.~\eqref{eq:f_G} numerically, and found
that the expressions in Table~\ref{table:GL1}
give a good estimate for these parameters.\cite{footnote4}
At the minima of the
free energy functional above we have 
$\delta\widetilde{\mathbf{\Lambda}} \propto \delta\widetilde{\mathbf{\mu}}$
and $\widetilde{\mathbf{\Delta}} \propto \sqrt{t}$. 
Therefore, the  expansion above contains all terms up to
$\mathcal{O}(t^2,\delta\widetilde{\mu} \, t, \delta\widetilde{\mu}^2)$.

The superfluid phase transition is driven by the 
term, $A_1(t,\widetilde \lambda)$, which changes sign at the 
$SU(3)$ point. All other coefficients are approximately constant 
close to the phase transition. 
The terms $\sim B_i$ describe the ferromagnetic order parameter, and
its response to the external "magnetic field", $\widetilde \mu$. 
The most interesting terms are the ones proportional to the
coefficients $C_i$: these describe the coupling between the SF order
parameter and the magnetization (or chemical potential differences), and 
they are responsible for the three-fold symmetric structure in
the central region of the phase diagram (see
Fig.~\ref{fig:MF_Phdiag_T=0.5Tc_alpha=0_0.5}). 
The terms $C_1$ and $C_2$ couple the
superfluid and magnetic order parameters, and produce
the density shift of the normal component at the onset of
superfluidity. 
Notice that all these terms are found to be 
proportional to the dimensionless 
particle-hole symmetry breaking parameter, $\widetilde \gamma$.

\begin{table}[t]
\begin{center}
\begin{tabular}{c|c}
\hline
\hline
parameter & approximate expression \\
\hline
\hline
$A_1$ &  $2.00 \; t +\dots $
\\
\hline
$A_2$ &  $0.40 - 1.20 \, t +\dots $
\\
\hline
$B_1$& $0.5000+1.000\,\widetilde{\lambda} +\dots $
\\
\hline
$B_2$ &   $ -1.000 \, \widetilde{\lambda}+\dots $
\\
\hline
$B_3$ & $ -1.000+\dots $
\\
\hline
$C_1$ & $ 1.25 \, \widetilde{\gamma}+\dots $
\\
\hline
$C_2$ & $-1.22 \, \widetilde{\gamma}+\dots $
\\
\hline
$C_3$& $ -0.62 \, \widetilde{\gamma}/{\widetilde\lambda}+\dots $
\\
\hline
\hline
\end{tabular}
\caption{ \label{table:GL1}
Approximate expressions of the Ginzburg-Landau coefficients 
in Eq.~\eqref{eq:GL_SU(3)_third_order}. The dimensionless parameters
are $\widetilde{\lambda}\equiv\rho_F \lambda$, $\widetilde{\gamma} \equiv
\gamma T_c$, and  $t \equiv (T-T_c)/T_c$.
}
\end{center}
\end{table}

While the third order expansion,
\eqref{eq:GL_SU(3)_third_order}  accounts for the central regions 
on the right panels of Fig.~\ref{fig:MF_Phdiag_T=0.5Tc_alpha=0_0.5},
it  does not
recover the sixfold symmetric structure  
that dominates the phase diagram at larger chemical potential
differences. 
This is obvious, since 
the terms $C_1, C_2$ and $C_3$ are odd under the 
particle-hole transformation, $\delta\widetilde{\mu} \leftrightarrow
-\delta\widetilde{\mu}, \delta\widetilde{\mathbf{\Lambda}} \leftrightarrow
-\delta\widetilde{\mathbf{\Lambda}}^*$, and are proportional to
$\gamma$,  while the hexagonal structure is 
even under particle-hole transformation, and already appears 
for $\gamma=0$.  The "hexagonal"  structure must therefore be  
 controlled by higher order terms, containing even degree polynomials of
$\delta\widetilde{\mu}$ and $\delta\widetilde{\mathbf{\Lambda}}$, coupled to
the SF order parameter. Unfortunately, the number of such terms is
huge, and  is next to impossible to determine 
all of them and their corresponding GL coefficients accurately. 
However, observing that
the ferromagnetic response is always small, we can just focus 
on the SF order parameter. At a formal level, this can be done
by minimizing the free energy functional $\tilde {f}_G$ in
$\delta\widetilde{\mathbf{\Lambda}}$ for any fixed  
$\delta\widetilde{\mathbf{\mu}}$ and
$\widetilde{\mathbf{\Delta}}$, and thus defining
$$
{\widehat{f}}_G(\delta\widetilde{\mathbf{\mu}},\delta\widetilde{\mathbf{\Delta}})
\equiv
\widetilde{f}_G(\delta\widetilde{\mathbf{\mu}},\delta\widetilde{\mathbf{\Delta}},\delta\widetilde{\mathbf{\Lambda}}_{\rm
min}(\delta\widetilde{\mathbf{\mu}},\delta\widetilde{\mathbf{\Delta}}))\;.
$$
The form of this GL functional is also dictated by symmetry, and it  can also be expanded in
$\delta\widetilde{\mathbf{\mu}}$ and
$\delta\widetilde{\mathbf{\Delta}}$. 
Up to $\mathcal{O}(t^2,\delta\widetilde{\mu}^2 \, t)$ it reads\cite{footnote4} 
\begin{align}
{\widehat{f}_G}&= a_1 \, \mathrm{Tr} (
\widetilde{\mathbf{\Delta}}\widetilde{\mathbf{\Delta}}^+ ) + a_2\,
\mathrm{Tr} (
(\widetilde{\mathbf{\Delta}}\widetilde{\mathbf{\Delta}}^+)^2)
\nonumber   \\ 
&+b \, \mathrm{Tr} (
\delta\widetilde{\mathbf{\mu}}\widetilde{\mathbf{\Delta}}\widetilde{\mathbf{\Delta}}^+)  
+ c_1\, \mathrm{Tr} ( \delta\widetilde{\mathbf{\mu}}^2
\widetilde{\mathbf{\Delta}}\widetilde{\mathbf{\Delta}}^+
) \label{eq:GL_SU(3)_fourth_order}\\ 
&+ c_2\, \mathrm{Tr} (
\delta\widetilde{\mathbf{\mu}}\widetilde{\mathbf{\Delta}}\delta\widetilde{\mathbf{\mu}}\widetilde{\mathbf{\Delta}}^+) 
+ \dots. \nonumber 
\end{align}
The approximate values of the numerically obtained coefficients are
enumerated in Table~\ref{table:GL2}.

Minimization of 
Eq.~\eqref{eq:GL_SU(3)_fourth_order}
 yields the correct
structure of the phase diagram
in the vicinity of the $SU(3)$ symmetric point, and accounts for  
the competition between the odd
($b,  \dots$) and even ($c_1,c_2, \dots$) order couplings. We
also checked that it determines correctly the absolute value
 of the SF order parameter in
the weak coupling regime $T_c/W < 0.1$
at temperatures $0.9 T_c < T < T_c$.  However, the locations of the 
\marci{triple} points at the interface of the threefold and approximately
sixfold symmetric structures in
Fig.~\ref{fig:MF_Phdiag_T=0.5Tc_alpha=0_0.5} are 
reproduced only  with an error of
about $50 \%$. Although this error is very large, it is also  natural, since on
the scale of this structure, the chemical potential difference is of
the order of $\delta\widetilde{\mu} \approx 0.2$. Therefore
$\delta\widetilde{\mu}$ cannot be considered as a small parameter here, 
and higher order terms in the expansion \eqref{eq:GL_SU(3)_fourth_order} 
shift the phase boundaries significantly. 

\begin{table}[t]
\begin{center}
\begin{tabular}{c|c}
\hline
\hline
parameter & approximate expression \\
\hline
\hline
$a_1$ & $2.0\;t +\dots $
\\
\hline
$a_2$ & $0.40 - 1.2 \, t +\dots $
\\
\hline
$b$ & $\left( 3.2 \, t - 0.083/\widetilde{\lambda}^2 \right) \,
\widetilde{\gamma} +\dots $
\\
\hline
$c_1$ & $0.125 - 0.29 \, \widetilde{\lambda} - 0.13 \, t+\dots $
\\
\hline
$c_2$ & $-0.115 + 0.27 \, \widetilde{\lambda} + 0.12 \, t +\dots $\\
\hline
\hline
\end{tabular}
\caption{\co \label{table:GL2}
Approximate expressions of the Ginzburg-Landau coefficients 
in Eq.~\eqref{eq:GL_SU(3)_fourth_order}. The dimensionless parameters
are $\widetilde{\lambda}\equiv\rho_F \lambda$, $\widetilde{\gamma} \equiv
\gamma T_c$, and  $t \equiv (T-T_c)/T_c$.
}
\end{center}
\end{table}

\section{Beyond mean-field}
\label{sec:beyond_MF}

In the  discussion presented so far we restricted ourselves to a 
mean-field approach, and neglected fluctuations. Fluctuations, however,
not only reduce somewhat the transition temperatures and fields, 
but they also change the universality class and thus the critical
exponents of the transition. 
In ordinary superfluids, such fluctuation effects are typically hard to
observe, however, in cold atomic systems 
one can reach the  strong coupling regime, and therefore 
a non-trivial critical behavior may be observable.\cite{KT_experiment}  

First, let us discuss the central $SU(3)$ symmetrical point 
of the phase diagram,  $\mu_x=\mu_y=0$. At this point only the 
first two terms of the GL action \eqref{eq:GL_SU(3)_fourth_order}
survive for an $SU(3)$ symmetrical interaction. 
These terms as well as the gradient term, 
${\mathrm Tr}\{\partial_{\bf r} {\mathbf{\Delta}}
\cdot \partial_{\bf r} {\mathbf{\Delta}}^+ \}$ have an 
increased O(6) symmetry with respect to $SU(3)$,\cite{footnote4} with 
the real and imaginary parts  of the independent 
components of  $\mathbf{\Delta}$  
forming a six component real vector. Since higher order terms
are irrelevant in the renormalization group (RG) sense, the $\mu_x=\mu_y=0$ transition 
is described by the O(6) critical theory. Thus the correlation length 
diverges as $\xi\sim |T-T_c|^{-\nu_{O(6)}}$, while the order parameter
scales as $\langle \mathbf{\Delta}\rangle\sim |T-T_c|^{\beta_{O(6)}}$.
\marci{For $d=3$ dimensions, the
critical exponents are known from $\epsilon$ 
expansions,\cite{O(n)ExponentsEpsilon} $1/n$ 
expansions,\cite{O(n)Exponents1N} as well as from 
high-temperature expansions,\cite{O(6)ExponentsHT} and Monte-Carlo
simulations,\cite{O(6)ExponentsMC} giving similar results,}
\be
\nu_{O(6)}^{3D} \approx  0.80,\quad \beta _{O(6)}^{3D} \approx 0.41\;.
\nonumber
\ee
In two dimensions, on the other hand, fluctuations suppress the phase
transition at the SU(3)-symmetrical point, $T_c^{2D}\to0$,\cite{Cardy} 
which thus becomes a quantum critical point.

For generic values of $\mu_x,\mu_y\ne0$, only one superfluid channel
dominates the phase transition, which is therefore 
described by the XY model. In $d=3$ dimension  the corresponding 
critical exponents are given by\cite{XYExponents} 
\bea
\nu_{XY}^{3D} &\approx&  0.67,\quad \beta _{XY}^{3D} \approx 0.35 ,
\eea
while in $d=2$ dimension the transition is of Kosterlitz-Thouless 
type.\cite{KTpaper}

Interesting critical behavior emerges in the vicinity of the 
bicritical lines of Fig.~\ref{fig:Intro_mudiff=0.25Tc_phdiag}. 
Along these lines, two components of the matrix ${\mathbf{\Delta}}$, 
e.g. $\Delta_{13}$ and $\Delta_{23}$ compete with each-other to 
form the superfluid. These can be grouped into a real four component 
vector,  $\varphi=(\mathrm{Re}
\Delta_{23},\mathrm{Im} \Delta_{23},\mathrm{Re}
\Delta_{13},\mathrm{Im} \Delta_{13})$. Fermion number conservation
implies that the effective action must be invariant under global phase
transformations, $\Delta_{ij}\to e^{i\phi_{ij}}\Delta_{ij}$, which
translates to an $O(2)\times O(2)$ symmetry in terms of the field 
$\varphi$. Up to fourth order, the 
most general effective Hamiltonian can be written as\cite{footnote6}
\begin{eqnarray}
H_{LW} = &\int \mathrm{d}^d x \, \left[ \frac{1}{2} (\nabla \varphi)^2
  + t_{+} \varphi^2 + t_{-} \varphi \mathbf{\Pi} \varphi
  \right. \label{eq:H_Landau_Wilson} \\ 
+& \left. u (\varphi^2)^2 + v (\varphi\mathbf{\Pi}\varphi)^2 + w
\varphi^2 (\varphi \mathbf{\Pi} \varphi) +\dots \right], \nonumber  
\end{eqnarray}
where the terms breaking the $O(4)$ symmetry were written in terms of  the
matrix 
\begin{equation}
\mathbf{\Pi} = \begin{pmatrix} 1 & & & \\ & 1 & & \\ & & -1 & \\ & & & -1 \end{pmatrix}.
\end{equation}
In the absence of the terms $t_{-}$ and $w$, this action has an additional
$\mathbb{Z}_2$ symmetry, $\Delta_{13}\leftrightarrow\Delta_{23}$,
leading to a  $O(2,2)=(O(2) \times O(2)) \rtimes \mathbb{Z}_2$ symmetry of the
free energy functional.  In the presence of particle-hole symmetry, one
  can show that at the boundary of the two superfluid phases 
the $\mathbb{Z}_2$ violating terms vanish: $t_-=w=0$. 
In general, however, the  simultaneous vanishing of $t_-$ and $w$ is
not guaranteed. Nevertheless, already leading order $\epsilon$
expansion indicates\cite{Epsilon_intro2} that the coupling $w$ is
irrelevant at the phase transition, $t_\pm\to 0$.
Thus the $\mathbb{Z}_2$ symmetry is apparently 
restored at the transition,  and the critical state  must be
described by the $O(2,2)$ symmetrical functional with $t_-,w\to0$.

\begin{figure}[t]
\includegraphics[width=8cm,clip=true]{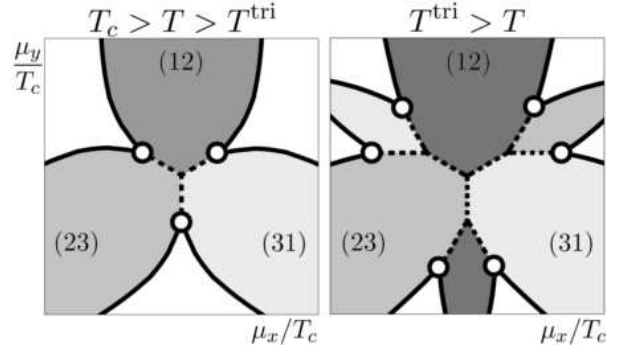}
\caption{\co  \marci{ Schematic picture of the position of the $O(2,2)$
points (empty circles) in case of linear DOS. 
At the temperature $T^{\mathrm{tri}}$ 
below which the triple points appear, 
from each $O(2,2)$ bicritical line (left) 
two new bicritical lines of the same universality class
branch out (right). The branching points are multicritical. 
SF-SF transitions are of first (dashed lines), whereas 
SF-N transitions are of second order (solid lines).} } 
\label{fig:Beyond_MF_O22Diagram}
\end{figure}

The  $O(2,2)$  functional \eqref{eq:H_Landau_Wilson} with  $t_-,w\to0$
thus describes the phase transition at all bicritical endpoints 
where two superfluid phases meet (white circles in
Fig.~\ref{fig:Beyond_MF_O22Diagram}). Notice that the structure of the
phase diagram changes close to  $T_c$, and the six $O(2,2)$
points, -- characteristic at lower temperatures, -- pairwise merge into three
  $O(2,2)$ points above a tricritical temperature, $T^\mathrm{tri}$,
as also shown in Fig.~\ref{fig:Beyond_MF_O22Diagram}).

The second order terms $t_{+}$ and $t_{-}$ trigger the SF-N and SF-SF
transitions, respectively, and scale as 
\begin{eqnarray}
t_{+} \propto & \, \delta\mu_{\parallel}, \\
t_{-} \propto & \, \delta\mu_{\perp},
\end{eqnarray}
for small chemical potential shifts parallel ($\delta\mu_\parallel$) and 
perpendicular ($\delta\mu_\perp$) to the SF-SF phase boundary.

The model \eqref{eq:H_Landau_Wilson} has been studied 
extensively,\cite{Epsilon_intro,Epsilon_intro2}
typically in the framework of the more general $n\cdot m$ component
models.\cite{DombGreen} Despite the extensive effort,  the
stability of its various fixed points is still debated. Systematic
$\epsilon$ expansion yields three non-trivial 
fixed points with $t^*_-=w^*=0$, which could potentially
describe the  critical state: (a) an $O(4)$ Heisenberg fixed point
with $u^*>0$ and $ v^* = w^* = 0$ (b)  a decoupled fixed 
point (DFP) ($u^* = v^*, w^* = 0$), where the two superfluid
components are described by two  independent XY theories, and 
(c)   a  mixed (or biconical) fixed point (MFP) with $u^* \ne v^*$ and
$w^* = 0$.

\begin{figure}[b]
\includegraphics[width=7cm,clip=true]{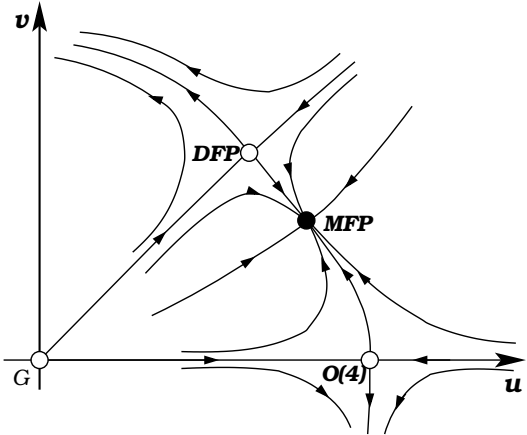}
\caption{\co  Schematic picture of the $\mathcal{O}(\epsilon^2)$ RG flows
  in the $(u,v)$ plane, with $t_{-}=w=0$, for $\epsilon<5/7$. For
  $\epsilon \rightarrow 1$ the fixed point structure changes, and 
 the $\epsilon$ expansion is inconclusive. } 
\label{fig:Beyond_MF_flows}
\end{figure}

For small values of $\epsilon=4-d$, $\epsilon$
expansion yields the picture shown in Fig.~\ref{fig:Beyond_MF_flows},
predicting that the mixed fixed point (MFP) describes the phase
transition along the $O(2,2)$ critical line. 
However, already in second order in $\epsilon$,\cite{Epsilon_second_order} 
the fixed point structure changes completely as one approaches 
the physical value, $\epsilon=1$, and even 
the results of six loop $\epsilon$ expansion remain completely 
inconclusive
regarding the stability of the fixed points.\cite{Epsilon_summary1} 
Non-perturbative arguments, on the other hand,  
seem to support  that the rather boring 
decoupled fixed point (DFP) describes the critical 
state.\cite{Epsilon_summary1,Epsilon_summary2,Epsilon_non_pert} 

\begin{figure}[t]
\includegraphics[width=8cm,clip=true]{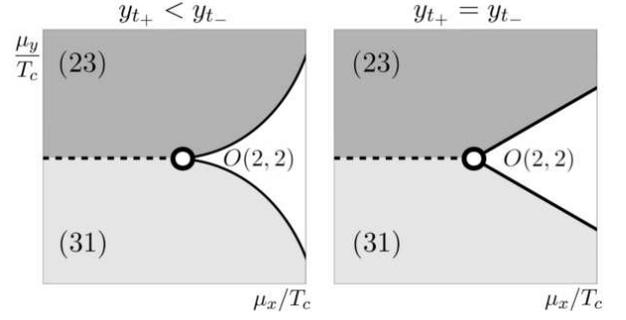}
\caption{\co  Schematic phase diagram of the vicinity of the $O(2,2)$
  \marci{bicritical} point (see Fig.~\ref{fig:Intro_schematic_phdiag}) in
  case of the mixed (left) and the decoupled (right) fixed point. In
  the former case fluctuations modify the SF-N phase boundary into
  curves with universal scaling.} 
\label{fig:Beyond_MF_O22}
\end{figure}

The universality class of the fixed point has considerable impact on
the phase diagram. The ratio of the critical exponents
$y_{\pm}$ associated with the terms $t_\pm$ 
determine e.g. the shape of the SF-N phase boundary in the vicinity 
of the bicritical point.
Standard  cross-over scaling arguments\cite{Cardy} lead to the 
conclusion, e.g., that the specific heat diverges in the vicinity of
the  SF-N  transition line as
\begin{equation}
c_v(t_{+})\propto \left|t_{+} - C(t_{-}) \right|^{-\alpha_{\rm XY}},
\end{equation}
where $\alpha_{\rm XY}$ denotes the specific heat exponent of the XY
model, and the phase boundary is determined by the function $C(t_{-})$
\begin{equation}
C(t_-) \propto |t_{-}|^{y_{+}/y_{-}}.
\end{equation}
Since the critical exponents $y_{{\pm}}$ 
are different for the two possible stable fixed points
even to first order in $\epsilon$,  
\bea
y_{{+}}^{\rm MFP} &=& 2-\epsilon/2+\dots, \quad 
y_{{-}}^{\rm MFP} = 2-\epsilon/6+\dots, 
\nonumber
\\
&&\quad y_{{\pm}}^{\rm DFP} =  2-\frac{2}{5}\epsilon+\dots, 
\eea
the shape of the phase boundary will be different in the two
cases. Notice  that since the DFP describes two independent $XY$
models, its exponents $y^{\rm DFP}_\pm$ will be equal to all orders in
$\epsilon$,  implying  that the SF-N boundaries start {\em linearly} at the
bicritical point. For the MFP, on the other hand, 
$y_{{+}}^{\rm MFP} <
y_{{-}}^{\rm MFP}$, and the SF-N boundary has a universal
exponent in the vicinity of the $O(2,2)$ point, as shown in
Fig.~\ref{fig:Beyond_MF_O22}. This difference in the
shape of the phase boundary provides a clear fingerprint 
of the universality class of the transition. 

The critical  exponent $\beta$ of the order parameter, $\langle \varphi
\rangle$  along the SF-SF phase boundary, is determined 
by the exponent $y_h$ of the "magnetic field" at the critical fixed
point,
$$
\beta = \frac{d-y_h}{y_{{+}}}\;.
$$
Since the magnetic field exponents get their first non-trivial
contribution in $\mathcal{O}(\epsilon^2)$ order, 
to leading order in $\epsilon$ we have
\bea
y_{{h}}^{\rm DFP}&=&3-\frac{\epsilon}{2}+\dots\;,
\nonumber \\
y_{{h}}^{\rm MFP}&=&3-\frac{\epsilon}{2}+\dots\;.
\eea
However, since $y_{{+}}^{\rm MFP} \neq y_{{+}}^{\rm DFP}$,
the exponents $\beta^{\rm DFP}$ and $\beta^{\rm MFP}$ 
turn out to be different  already to first order in 
$\epsilon$,
\be
\beta^{\rm DFP}=\frac 1 2 - \frac {3}{20} \epsilon +\dots\;,\quad
\beta^{\rm MFP}=\frac 1 2 - \frac {\epsilon}{8} + \dots\;.
\ee

\section{Experimental relevance} \label{sec:ExpRelev}

Currently maybe  $\left.^6 \mathrm{Li} \right.$ ultracold gases
provide the most promising  perspective for the realization of 
three component superfluidity. For high magnetic fields, 
the s-wave scattering lengths between the three lowest hyperfine 
states approach the spin-triplet scattering length, $a_{12}\approx a_{23}\approx
a_{31}\approx -2140 \, a_0$, with $a_0$ the Bohr
radius.\cite{Experimental_6Li_scattering_lengths} 
At fields of $\sim 2000 \;{\rm G}$, for example, 
the scattering lengths all deviate less than $2\%$ from
their average value\cite{Experimental_6Li_scattering_lengths}. 
It has been proposed theoretically that 
this deviation can further be decreased 
using radio frequency and microwave
fields,\cite{Experimental_Make_sc_lengths_equal}  
down to $0.1\,\%$, and thereby a strongly
attractive system can be realized with almost perfect $SU(3)$ symmetry
in this high field regime.

Although three-body loss is a major obstacle in three component
experiments, recent experiments showed that decay rates
tend to decrease at high
fields in $^6{\rm Li}$ systems, and indeed, 
Fermi degeneracy has successfully been realized in this three component
system.\cite{OHara}
%
A  $^6\mathrm{Li}$ experiment  on a system
of Fermi energy $T_F=1 \mu\mathrm{K}$ and 
without optical lattice would correspond to the parameters
$\lambda_{\alpha \ne \beta} \rho_F\approx 0.11, \gamma W \approx 0.18$ 
and $T_c / W \approx 0.01$.\cite{Unpublished} This system would thus be in the regime of 
weak interactions, studied here. However,   such a small critical 
temperature is currently unreachable. 
Application of an optical lattice can, however, easily bring the 
system into the regime of strong interactions, where $SU(3)$
superfluidity may be  accessible. Though our calculations do not
apply for strong interactions, we believe that, similar to 
the $SU(2)$ case,\cite{Sarma,SfExperiment2,FFLOExp1,DensityJumpExp} 
the major features of our  phase diagram are robust,
and should carry over to the strongly interacting case.

\begin{figure}[h]
\includegraphics[width=8cm,clip=true]{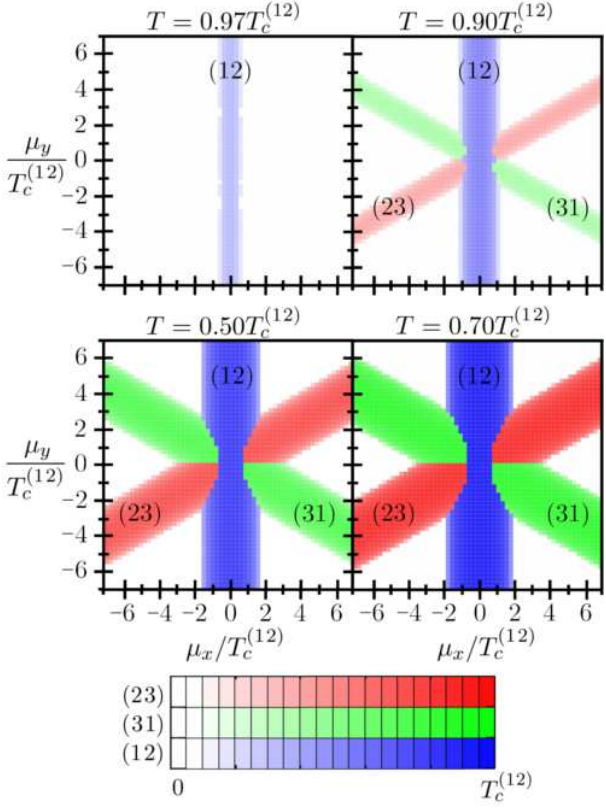}
\caption{\co  Phase diagram with only approximately equal interaction
  strengths. The rate of the critical temperatures in the respective
  channels are
  $T_c^{(23)}/T_c^{(12)}=T_c^{(31)}/T_c^{(12)}=0.95$. The SF phase
  in the channel of strongest interaction repels the other two
  phases from the central region of the phase diagram. Parameters at
  the $\mu_x=\mu_y=0$ point:
  $(\lambda_{12},\lambda_{23},\lambda_{31})\rho_F=(0.1057,0.1046,0.1046),
  T_c^{(12)} W=0.01, \gamma W = 0, \xi_F=0$.  } 
\label{fig:Exp_relev_mudiff=7Tc12_alpha=0DOS_Tc=0.0100_0.0095_0.0095_lambda=0.1057_0.10455_0.10455} 
\end{figure}

So far we assumed a perfectly $SU(3)$ symmetrical interaction in our
calculations. The phase diagram is, however, somewhat modified   
if the  the scattering lengths are only
approximately equal.\cite{UnequalInteractionStrengths,Catelani2008} In
Fig.~\ref{fig:Exp_relev_mudiff=7Tc12_alpha=0DOS_Tc=0.0100_0.0095_0.0095_lambda=0.1057_0.10455_0.10455}
we present a phase diagram for the case 
where we have set the ratio of critical temperatures in the
different channels to be  $T_c^{(23)}/T_c^{(12)}
=T_c^{(31)}/T_c^{(12)}=0.95$. For $T_c^{(ij)}/W\sim 0.01$ 
this would correspond to 
 a $\sim 1\%$ asymmetry of the scattering lengths.
At temperatures $T_c^{(12)} > T > T_c^{(23)}=T_c^{(31)}$ the SF phase is
formed only in the $(12)$ channel. The star-like shape of the phase
diagram is preserved at lower temperatures, 
however, the interaction asymmetry destroys the
sixfold symmetry of the central region of the phase diagram, including 
the $O(6)$ critical point: the phase $(12)$ dominates this  
central region and expels the other two SF phases.
Thus the shape
of this region depends rather sensitively on the interaction
asymmetry, and fine tuning of the scattering lengths 
(by using RF and MW fields,\cite{Experimental_Make_sc_lengths_equal}
e.g.) may be needed  to realize an $SU(3)$ symmetric superfluid.  

\section{Conclusions}

In this paper, we studied the phase diagram and the interplay of
fermionic and superfluid order parameters in a three component
fermionic mixture. We mostly focused on the case of $SU(3)$
symmetrical interactions, and studied 
 the weak coupling regime, where 
the critical temperature is much smaller than the Fermi energy of the
atoms, $T_c<E_F$. We combined two complementary mean
field methods (Gaussian variational method, and equation of motion
techniques) to study how a chemical potential
imbalance polarizes the atomic cloud and modifies/destroys superfluid order.
Though the phase diagram of the three component system is naturally 
 much richer than that of the two component mixture,\cite{Sarma} 
there are some similarities: 
large chemical potential imbalances
($|\mu_i-\mu_j|\gg T_c$ for all $i\ne j$), for example,  
destroy superfluid (SF) order, similar to two component mixtures.  
The corresponding SF-normal
transition is of second order at higher temperatures, while it 
becomes of first order below the Sarma
temperature.

\begin{figure}[t]
\includegraphics[width=4cm,clip=true]{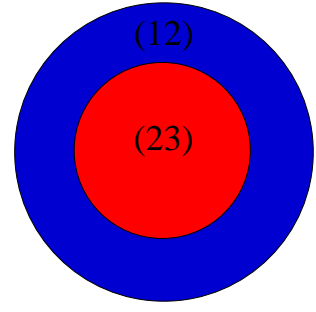}  
\caption{\co  \marci{Possible trap configurations for total 
atom numbers $N_3>N_2>N_1$.} } 
\label{fig:ConclusionsTraps}
\end{figure}

The superfluid phase is, on the other hand, much richer than in the
two component case. SF order can form in channels $(12)$, $(23)$, and 
$(31)$, and the chemical potential driven transitions between these
 superfluid phases are of first order. In a real experiment, 
where fermion numbers are approximately conserved for each component, 
such first order transitions would appear as segregation 
of different SF phases, and domain formation.\cite{Rapp}
Experimentally, these domains would probably 
 appear as a shell  structure, sketched in 
Fig.~\ref{fig:ConclusionsTraps}. For $N_3>N_2>N_1$, e.g., one expects that
in the center of the trap a $(32)$ superfluid forms, however, 
approaching the external region of the trap $T_c$ decreases, and 
the $(31)$ superfluid state becomes more stable.

As a rule of thumb, SF order tends to 
form in the channel of the smallest chemical potential 
difference. This simple rule determines the overall structure of the phase
diagram (see Fig~\ref{fig:MF_Phdiag_T=0.5Tc_alpha=0_0.5}). 
However, unlike the two component case, for three component mixtures
a non-trivial coupling between   magnetic and SF order is
also allowed.\cite{Demler,Rapp} This interesting coupling --- the strength 
of which is regulated by particle-hole symmetry breaking, $\gamma\sim
\varrho'(E_F)/\varrho(E_F)\sim 1/E_F$ --- leads to a peculiar  
triangular structure in the central region of the phase
diagram, $\mu_i\approx \mu$, in agreement with
the predictions of Ref.~\onlinecite{Demler} (though with opposite
orientation, see Fig~\ref{fig:MF_Phdiag_T=0.5Tc_alpha=0_0.5}).  
However, the relative
size of this central  region is apparently  proportional to 
$\sim \sqrt{\gamma T_c}$; therefore, for weak and intermediate
couplings, the triangular structure appears only in the close vicinity
of the $SU(3)$   symmetrical point, $\mu_i\equiv \mu$. For very strong
attractive interactions, $T_c\approx E_F\sim W$, on the other hand,
the central (triangular) region must get more extended, and may become
observable. 

We also constructed the Ginzburg-Landau functionals describing the
three component mixture, and determined  the temperature and asymmetry
($\gamma$) dependence of the various coefficients. We have shown that,
to capture the termination of the
central triangular region, one needs to go beyond the expansion of
Ref.~\onlinecite{Demler}, and higher order terms need be incorporated
in the functionals.

As discussed in Sec.~\ref{sec:beyond_MF}, fluctuations modify somewhat
the mean-field picture. The temperature-driven
phase transition for generic (unequal) chemical  potential values is
typically described by the XY model and its critical exponents. 
However, for certain special 
chemical potentials, the competition between various superfluid orders
may lead to interesting critical behavior. For $\mu_i\equiv\mu $ and  an
$SU(3)$ symmetrical interaction, e.g., the normal-SF transition belongs
to the $O(6)$  universality class, and is characterized by the
corresponding exponents. Along the critical lines separating the 
three phases, $(12)$, $(23)$, and $(31)$, on the other hand, an
interesting $O(2,2)$ critical behavior may emerge (see our discussion in
Sec.~\ref{sec:beyond_MF}). The shape of the phase diagram in the
vicinity of these special lines is then determined by the
corresponding universal cross-over exponents.  We emphasize that -- 
while it is very
difficult to observe it in the weak coupling regime --  a non-trivial 
critical behavior could be observable in the strong coupling regime,
often reached in cold atom experiments.

Finally, we studied the fragility of the $SU(3)$ physics, i.e.,
the sensitivity of these results and the phase diagram
to the symmetry of interaction. We have shown that 
already a small difference in the scattering lengths can 
substantially distort the $SU(3)$ phase diagram, and the SF phase of
the channel with the strongest interaction may suppress and 
mask the  $SU(3)$ symmetrical ($O(6)$) critical regime. 
These results agree with those obtained in 
Ref.~\onlinecite{Catelani2008}. Here, however, in contrast to 
Ref.~\onlinecite{Catelani2008}, we focused 
on the consequences of SU(3) symmetry (rather than on the consequences
of its violation), and the effects of the coupling between
ferromagnetic and superfluid order parameters,  neglected 
in  Ref.~\onlinecite{Catelani2008}. In addition, we also 
discussed the role of fluctuations and the structure of the emerging critical
states and multicritical lines.
Our results as well as those of Ref.~\onlinecite{Catelani2008} 
indicate that in
experimental realizations, to observe the $SU(3)$ 
physics, one should use systems with almost perfectly
symmetrical interactions, similar to Yb\cite{YbSymmetric},
or one should use some tricks to make all scattering lengths 
equal as much as possible.\cite{Experimental_Make_sc_lengths_equal}
Moreover, one should possibly stay in the strong coupling regime,
$T_c\sim W$, where the impact of a small asymmetry in the interaction is not
exponentially large.

\section{Acknowledgment}

We would like to thank Eugene Demler, Gil Refael, and Walter
Hofstetter for enlightening discussions. This research has
been supported by the   Hungarian research funds 
OTKA and NKTH under Grant Nos.~K73361 and CNK80991.
G.Z. acknowledges support  from the Humboldt Foundation and 
the DFG.

\appendix

\section{Exact Ward identities} \label{sec:Ward1}

In this Appendix, by  making use of the global $SU(3)$ invariance 
of the functional measure, we derive
exact Ward identities that give constraints on the possible values of the order
parameters and densities, Eqs.~(\ref{eq:Delta_def}-\ref{eq:d_def}). 
 
Consider the partition function $\mathcal{Z}$, defined in Eq.~\eqref{eq:Z_def}. 
For the current calculation we rewrite the
action Eqs.~(\ref{eq:S_0_def},\ref{eq:S_int_def}) in the form 
\begin{align}
S_0(\hat{\mu}) &=\sum_{\alpha\beta} \int \mathrm{d}x\, 
  \overline{\psi}_\alpha \left( (\partial_\tau + \mathcal{H}_0) \delta_{\alpha\beta} 
  - \hat{\mu}_{\alpha\beta} \right) \psi_\beta, \\
S_\mathrm{int}(\Gamma) &=-\sum_{\alpha\beta\gamma\delta}
  \Gamma_{\alpha\beta\gamma\delta} \int \mathrm{d}x\, 
  \overline{\psi}_\alpha \overline{\psi}_\beta \psi_\gamma \psi_\delta\;. 
  \label{eq:S_int_Gamma_def}
\end{align}
by introducing $\hat{\mu}_{\alpha\beta}=\mu_\alpha \, \delta_{\alpha\beta}$ 
and $\Gamma_{\alpha\beta\gamma\delta}= \frac{1}{2} \lambda_{\alpha\beta}\left( \delta_{\alpha\delta}\delta_{\beta\gamma} - \delta_{\alpha\gamma}\delta_{\beta\delta}\right)$.
An $SU(3)$ transformation of the fields 
$\psi_\alpha (x) \rightarrow \sum_\beta U_{\alpha\beta} \psi_\beta (x)$
translates to the transformation of $\hat\mu$ and $\Gamma$ in the functional integral.
Expressing $\mathbf{U}=\exp(i \sum_{a=1}^8 \eta^a \mathbf{T}^a)$ with the Gell-Mann matrices
$\mathbf{T}^a$, we find
\begin{eqnarray}
\left. \frac{\partial}{\partial\eta^a} \hat{\mu}_{\alpha\beta} (\eta) \right|_{\eta^a=0} & = i \sum_{\gamma} \left( \hat{\mu}_{\alpha\gamma} T^a_{\gamma\beta} - T^a_{\alpha\gamma} \hat{\mu}_{\gamma\beta} \right), \label{eq:Ward_Upsilon_eq} \\
\left. \frac{\partial}{\partial\eta^a} \Gamma_{\alpha\beta\gamma\delta} (\eta) \right|_{\eta^a=0} & = 2i \left( \lambda_{\alpha\beta} - \lambda_{\gamma\delta} \right) \delta_{\alpha\delta} \, T^a_{\beta\gamma}\;. \label{eq:Ward_Gamma_eq}
\end{eqnarray}

The invariance of the functional integral with respect to global $SU(3)$ transformations,
$\left. \frac{\partial \mathcal{Z}}{\partial \eta^a} \right|_{\eta_a=0}=0$,
leads to the Ward identity
\be
\left( \mu_\alpha - \mu_\beta \right) \frac{\partial
  \mathrm {ln}\mathcal{Z}}{\partial \hat{\mu}_{\alpha\beta}} = 2 \sum_\gamma \left(
\lambda_{\beta\gamma} - \lambda_{\alpha\gamma} \right) \frac{\partial
  \mathrm {ln}\mathcal{Z}}{\partial \Gamma_{\gamma \alpha \beta \gamma}}, 
\ee
for any $\alpha$ and $\beta$, from which Eq.~\eqref{Ward} follows.

\section{Ward identities in the Gaussian approximation}
\label{sec:Ward2}
Here, we derive approximate Ward identities, 
similar to those in Appendix~\ref{sec:Ward1}, that
 hold in the Gaussian approximation.  
As explained in Appendix~\ref{App:Saddle_point}, we can assume that 
the inverse propagator in the definition of the partition function 
$\mathcal{Z_D}$, Eq.~\eqref{eq:Z_D_def},
is local,
\be
\mathcal{Z}_\mathcal{D} = \int \mathfrak{D}\mathbf{\overline{\psi}} \, 
\mathfrak{D}\mathbf{\psi} \, 
e^{\frac{1}{2} \int \mathrm{d}1\, \overline{\phi}(1) \mathbf{D}^{-1}(1) \phi(1)},
\ee
where $\mathbf{D}^{-1}$ is defined in Eq.~\eqref{eq:Nambu_propagator}. 

An $SU(3)$ transformation of the fields 
$\psi_\alpha (x) \rightarrow \sum_\beta U_{\alpha\beta} \psi_\beta (x)$
translates to the transformation of order parameters,
\begin{align}
\mathbf{\Lambda} & \mapsto \mathbf{U} \mathbf{\Lambda} \mathbf{U}^+, \\
\mathbf{\Delta} & \mapsto \mathbf{U} \mathbf{\Delta} \mathbf{U}^\mathrm{T},
\end{align}
see Eqs.~(\ref{n_trans},\ref{d_trans}). Using
the invariance of the partition function with 
respect to these global $SU(3)$ transformations, we get the following 
constraints on the densities,
\be
\mathrm{Tr} \left( 
\begin{pmatrix} 
\mathbf{\Gamma}_\mathbf{\Lambda}^a & \mathbf{\Gamma}_\mathbf{\Delta}^a \\
-\mathbf{\Gamma}_\mathbf{\Delta}^{a +} & \mathbf{\Gamma}_\mathbf{\Lambda}^{a *}
\end{pmatrix}
\begin{pmatrix}
-\mathbf{n}^* & \mathbf{d} \\
\mathbf{d}^+ & \mathbf{n}
\end{pmatrix}
\right) = 0\;,
\label{eq:WardGaussian}
\ee
with $\mathbf{\Gamma}_\mathbf{\Lambda}^a = \left[ \mathbf{\Lambda},\mathbf{T}^a \right]$ and $\mathbf{\Gamma}_\mathbf{\Delta}^a=2 \left( \mathbf{T}^a \mathbf{\Delta} + 
\mathbf{\Delta} \mathbf{T}^{a *} \right)$.\cite{footnote7} Here the matrices 
$\mathbf{T}^a$, $a=1, \dots, 8$, are the Gell-Mann matrices.

In case of $SU(3)$ symmetric interactions,
at the solutions of the EOM equations,
Eqs.~(\ref{eq:Delta_def},\ref{eq:Lambda_def},\ref{eq:gapeq}), 
this equation simplifies to the same form
as the exact Ward identity, Eq.~\eqref{Ward},
\be
\left( \mu_\alpha - \mu_\beta \right) n_{\alpha\beta} = 0\;.
\ee
Therefore, when neither two of the chemical potentials are equal, 
the matrix of densities $\mathbf{n}$ and that of renormalized 
chemical potentials $\mathbf{\Lambda}$ are both diagonal 
(see Eq.~\eqref{eq:Lambda_def}).

\section{Saddle point equation in the Gaussian approximation} \label{App:Saddle_point}

In this Appendix, starting from the saddle point equation, Eq.~\eqref{eq:saddle_point}, 
we derive the saddle point form of the propagator $\mathcal{D}$
in the Gaussian approximation, Eqs.~(\ref{eq:D=D},\ref{eq:Nambu_propagator}). 
We will use the notations of 
Section~\ref{sec:Gaussian_var}.

First, we fix the arbitrariness in the form of $\mathcal{D}^{-1}$ 
in the definition of $S_\mathcal{D}$, Eq.~\eqref{eq:S_D_def}. 
We split $\mathcal{D}^{-1}$ into $3 \times 3$ matrices
\be
\mathcal{D}^{-1}(1,2)= 
\begin{pmatrix} 
\mathbf{\Gamma}_A(x_1,x_2) &
\mathbf{\Gamma}_B(x_1,x_2) \\
\mathbf{\Gamma}_C(x_1,x_2) &
\mathbf{\Gamma}_D(x_1,x_2)
\end{pmatrix}.
\ee
It is easy to see, that because of the anticommutation of the fields 
$\overline{\psi}_\alpha$ and $\psi_\alpha$, 
modifications of $\mathcal{D}^{-1}$ that leave 
$\mathbf{\Gamma}_A(x_1,x_2)-\mathbf{\Gamma}_D^T(x_2,x_1)$, 
$\mathbf{\Gamma}_B(x_1,x_2)-\mathbf{\Gamma}_B^T(x_2,x_1)$ and 
$\mathbf{\Gamma}_C(x_1,x_2)-\mathbf{\Gamma}_C^T(x_2,x_1)$ invariant, 
will not change $S_\mathcal{D}$. Therefore we may assume
that $\mathcal{D}^{-1}$ has the symplectic symmetry
\be
\begin{pmatrix}
\mathbf{0} & \mathbf{1} \\
\mathbf{1} & \mathbf{0}
\end{pmatrix}
\, \mathcal{D}^{-1} (x_1,x_2) \, 
\begin{pmatrix}
\mathbf{0} & \mathbf{1} \\
\mathbf{1} & \mathbf{0}
\end{pmatrix}
= -\left(\mathcal{D}^{-1}\right)^\mathrm{T} (x_2,x_1). 
\label{eq:Symplectic_Grassmann}
\ee


The saddle point equation, 
Eq.~\eqref{eq:saddle_point}, gives very strong constraints 
on the form of $\mathcal{D}$. In particular, 
it is equivalent to the EOM self-consistency 
equation of Section~\ref{sec:EOM_technique}. To see this, we use the definition
Eq.~\eqref{eq:Feynmans_ineq} to  rewrite Eq.~\eqref{eq:saddle_point}
 in the form
\be
\frac{1}{\mathcal{Z}_\mathcal{D}} \, 
  \frac{\delta \mathcal{Z}_\mathcal{D}}{\delta \mathcal{D}(1,2)} = 
  \frac{\delta \langle S-S_\mathcal{D} \rangle_\mathcal{D}}
  {\delta \mathcal{D}(1,2)}. \label{eq:variation1}
\ee
The calculation of the left hand side of this equation is straightforward. Using
only the definition of $\mathcal{Z_D}$ (see Eq.~\eqref{eq:Z_D_def}), and 
Eq.~\eqref{eq:Grassmann_propagator}, we get
\be
\frac{1}{\mathcal{Z}_\mathcal{D}} \, 
  \frac{\delta \mathcal{Z}_\mathcal{D}}{\delta \mathcal{D}(1,2)}= 
  -\frac{1}{2} \mathcal{D}^{-1}(2,1).
\ee
To evaluate the right hand side of Eq.~\eqref{eq:variation1}, 
omitting a constant term, we can write
\be
\langle S-S_\mathcal{D} \rangle_\mathcal{D} = 
  -\frac{1}{2} \int \mathrm{d}1 \,\mathrm{d}2 \,  
  \mathcal{D}_0^{-1}(1,2) \mathcal{D}(2,1) + 
  \langle S_\mathrm{int} \rangle_\mathcal{D}.
\ee
Then, it is easy to see that, the saddle point equation is equivalent to
\be
\mathcal{D}^{-1}(1,2) = \mathcal{D}_0^{-1}(1,2)-
  2 \, \frac{\delta \langle S_{\mathrm{int}} \rangle_\mathcal{D}}
  {\delta \mathcal{D}(2,1)}.
\ee
Expanding $\langle S_\mathrm{int} \rangle_\mathcal{D}$ using 
Wick's theorem gives a product of equal time propagators, 
whose variation according to the propagator matrix $\mathcal{D}$ 
can be straightforwardly calculated. We get the desired formulas, 
Eqs.~(\ref{eq:D=D},\ref{eq:Nambu_propagator}), with the order 
parameters $\mathbf{\Lambda}$ and $\mathbf{\Delta}$ satisfying 
the EOM self-consistency equations, 
Eqs.~(\ref{eq:Delta_def},\ref{eq:Lambda_def}), and \eqref{eq:gapeq}. 
This means, that the EOM method is \textit{consistent} with the Gaussian 
variational approach.

\section{Calculation of the Gaussian approximation to the free energy} 
\label{App:Calc_free_energy}

In the following we calculate the Gaussian approximation of the 
free energy, Eq.\eqref{eq:F_G_operator_formalism}. We first introduce the
Fourier components $\psi_\alpha (\mathbf{r}) = \frac{1}{\sqrt{\Omega}}
\sum_\mathbf{k} e^{i \mathbf{k r}} a_{\alpha\mathbf{k}}$, obeying the
anti-commutation relations 
$\lbrace a^\dagger_{\mathbf{k}\alpha},a_{\mathbf{k}' \beta} \rbrace = 
\delta_{\alpha\beta} \, \delta_{\mathbf{k}\mathbf{k}'}$, 
where $\Omega$ denotes the volume.
With these, the Hamiltonian, Eq.~\eqref{eq:H_D_def}, takes on the form
\be
H_\mathcal{D}= \frac{1}{2} \sum_\mathbf{k}\left\{ 
  \left( \mathbf{a}^\dagger_\mathbf{k}, \mathbf{a}_{-\mathbf{k}} \right)
  \mathbf{B}(\xi_\mathbf{k})
  \begin{pmatrix}
    \mathbf{a}_\mathbf{k} \\ \mathbf{a}^\dagger_{-\mathbf{k}}
  \end{pmatrix} 
  + \mathrm{Tr} \left( \xi_\mathbf{k} - \mathbf{\Lambda} \right)\right\},
\ee
with $\mathbf{B}(\xi)$ defined in Eq.~\eqref{eq:B_def}, and the last
term originating from normal ordering.

From the above form, the calculation of $\mathcal{Z}_\mathcal{D} =
\mathrm{Tr}e^{-\beta H_\mathcal{D}}$ 
is straightforward, though some care is needed to avoid double counting 
in momentum space. Note that, because of the symplectic symmetry, 
Eq.~\eqref{eq:B_symplectic}, and Hermiticity of the matrix $\mathbf{B}(\xi)$,
its eigenvalues are real and come in pairs.
To each eigenvalue $\eta(\xi)$ there is another eigenvalue 
$-\eta\left(\xi\right)$. Using this property, $\log\mathcal{Z}_\mathcal{D}$ 
simplifies to
\begin{align}
\log \mathcal{Z}_\mathcal{D} =& \frac{1}{2} \sum_\mathbf{k} 
  \mathrm{Tr} \log \left( 2 \cosh \left( 
  \frac{\beta}{2} \mathbf{B}\left( \xi_\mathbf{k} 
  \right) \right) \right) \label{eq:logZ_D} \\ 
-&\frac{\beta}{2} \sum_\mathbf{k} \mathrm{Tr} \left( 
  \xi_\mathbf{k} - \mathbf{\Lambda} 
  \right). \nonumber
\end{align}
The calculation of $\langle H-H_\mathcal{D} \rangle_\mathcal{D}$ 
is also straightforward using Wick's theorem. One finds
\begin{align}
\frac{1}{\Omega} \langle H-H_\mathcal{D} \rangle_\mathcal{D} =& 
  \sum_{\alpha\beta} \lambda_{\alpha\beta} 
  (\left| n_{\alpha\beta} \right|^2-n_{\alpha\alpha}n_{\beta\beta} - 
  \left| d_{\alpha\beta} \right|^2 ) \nonumber \\ 
+& \sum_{\alpha\beta} \left( \Lambda_{\alpha\beta} -
  \mu_\alpha \delta_{\alpha\beta} \right) n_{\alpha\beta} \label{eq:H-H_D} \\ 
+& \sum_{\alpha\beta} \Delta_{\alpha\beta} d^*_{\alpha\beta} + 
  \Delta^*_{\alpha\beta} d_{\alpha\beta}. \nonumber
\end{align}
Thus, using
Eqs.~(\ref{eq:logZ_D},\ref{eq:H-H_D}), we get the result Eq.~\eqref{eq:f_G}
for the Gaussian approximation of the free energy density .

In order to evaluate Eq.~\eqref{eq:H-H_D}, the densities and
anomalous densities, $\mathbf{n}$ and $\mathbf{d}$, also have to be determined. 
These can be easily calculated from the variations of \eqref{eq:logZ_D} 
with respect to $\mathbf{\Lambda}$ and $\mathbf{\Delta}$,  leading to 
the same equation, Eq.~\eqref{eq:gapeq}, 
as the EOM self-consistency equations.

\section{Particle-hole transformation}
\label{appendix_ph}

Particle-hole symmetry introduces a $\mathbb{Z}_2$ 
symmetry of the mean-field phase diagram, when
the band is half-filled, the DOS is particle-hole symmetric
($\rho(\xi) = \rho(-\xi)$),
and the interaction has $SU(3)$ symmetry
($\lambda_{\alpha\ne\beta} = \lambda$). 
This symmetry together with the permutation
symmetry of the fermion species makes the phase diagram six-fold symmetric,
see Fig.~\ref{fig:Intro_schematic_phdiag}. 

In this Appendix we calculate the 
effect of the particle-hole transformation
\be
\Psi_\alpha(x) \longleftrightarrow \Psi^\dagger_\alpha(x)
\ee
on the order parameters 
$\mathbf{\Lambda}$ and $\mathbf{\Delta}$.
This transformation leaves the interaction invariant, whereas
it modifies the bare chemical potentials and the single particle
energies as
\bea
\mathcal{H}_0 & \rightarrow & -\mathcal{H}_0, \\
\mu_\alpha & \rightarrow & -\mu_\alpha - 4 \lambda \, n_{\mathrm{max}},
\eea
where $n_\mathrm{max}=\int_{-W}^W \mathrm{d}\xi \, \rho(\xi)$ is the 
density of the completely filled band. The
bare chemical potentials remain unchanged on
the mean-field level at
\be
\mu_\mathrm{half} = - 2 \lambda \, n_\mathrm{max},
\ee
which is precisely the condition for the band being half-filled 
(see Eq.~\eqref{eq:Lambda_def}).

In order to investigate the inversion symmetry of the phase diagram,
consider two Hamiltonians with opposite differences in bare chemical
potentials from half-filling,
\bea
H^{(1)} &\equiv& H(\mathcal{H}_0,\mu_\mathrm{half}+\delta\mu_\alpha,
                   \lambda,\Psi^\dagger_\alpha,\Psi_\alpha), 
                   \label{eq:H(1)_def} \\
H^{(2)} &\equiv& H(\mathcal{H}_0,\mu_\mathrm{half}-\delta\mu_\alpha,
                   \lambda,\Psi^\dagger_\alpha,\Psi_\alpha),
                   \label{eq:H(2)_def}
\eea
as defined in Eq.~\eqref{eq:H_0_def}.
A particle-hole transformation of $H^{(2)}$ leads to
the equation
\be
H^{(2)} = H(-\mathcal{H}_0,\mu_\mathrm{half}+\delta\mu_\alpha,
              \lambda,\widetilde{\Psi}^\dagger_\alpha,\widetilde{\Psi}_\alpha)
        \equiv H^{(3)}, \label{eq:H(3)_def}
\ee
where $\widetilde{\Psi}_\alpha=\Psi^\dagger_\alpha$.
Accordingly, the densities in the original and the particle-hole
transformed system can be connected as
\bea
n_{\alpha\beta}^{(3)} &\equiv& \langle \widetilde{\Psi}^\dagger_\alpha(x) \widetilde{\Psi}_\beta(x) \rangle_{(3)}
                       =  -n_{\alpha\beta}^{(2)*} + n_\mathrm{max}, \\
d_{\alpha\beta}^{(3)} &\equiv& \langle \widetilde{\Psi}_\alpha(x) \widetilde{\Psi}_\beta(x) \rangle_{(3)}
                       =  -d_{\alpha\beta}^{(2)*}.
\eea
Then, it is also straightforward to show from the definitions 
Eqs.~(\ref{eq:Delta_def},\ref{eq:Lambda_def}), that the relation
between the order parameters are
\be
\mathbf{\Lambda}^{(2)} = -\mathbf{\Lambda}^{(3)*}, 
\hspace{12 pt}
\mathbf{\Delta }^{(2)} = -\mathbf{\Delta }^{(3)*}.
\label{eq:Lambda(2)=-Lambda(3),eq:Delta(2)=-Delta(3)}
\ee

Looking at their definitions, we see that the only difference
between $H^{(2)}$ and $H^{(3)}$ is in the sign of $\mathcal{H}_0$.
However, if the DOS is electron-hole symmetric,
\be
\rho(\xi) = \rho(-\xi),
\ee
then all of the EOM self-consistency equations 
Eqs.~(\ref{eq:Delta_def},\ref{eq:Lambda_def},\ref{eq:gapeq}), 
and the mean-field free energy Eqs.~(\ref{eq:gapeq},\ref{eq:f_G}) 
are identical in the two systems. Therefore, the set of the 
possible mean-field configurations have to be the same
($\mathbf{\Lambda}^{(1)} = \mathbf{\Lambda}^{(3)}$, 
 $\mathbf{\Delta }^{(1)} = \mathbf{\Delta }^{(3)}$). Putting
this, and Eq.~(\ref{eq:Lambda(2)=-Lambda(3),eq:Delta(2)=-Delta(3)})
together, we obtain the desired equations
\bea
\mathbf{\Lambda}(\mu_\mathrm{half}+\delta\mu_\alpha) &=& 
    -\mathbf{\Lambda}^{*}(\mu_\mathrm{half}-\delta\mu_\alpha), \\
\mathbf{\Delta }(\mu_\mathrm{half}+\delta\mu_\alpha) &=& 
    -\mathbf{\Delta }^{*}(\mu_\mathrm{half}-\delta\mu_\alpha),
\eea
connecting order parameters at opposite $\delta\mu_\alpha$ values, 
with the other parameters of the system unchanged.

We remark, that in the special case when $\delta\mu_1+\delta\mu_2+\delta\mu_3=0$, 
the particle-hole symmetry connects the points of the same $(\mu_x,\mu_y)$ plane,
and the mean-field phase diagram has an inversion symmetry. Away from this plane
the inversion symmetry is only approximate, due to logarithmic corrections
to the values of the order parameters, coming from the asymmetric cut-off.


\begin{thebibliography}{}
\bibitem{footnoteInteractionStrengths} Note that our convention for the interaction strength
    differs from the usual convention by a factor of 1/2. This difference also affects 
    the definitions of the order parameters later, see Eqs.~(\ref{eq:Delta_def},\ref{eq:Lambda_def}).
\bibitem{footnote0}
Notice that the chemical potentials 
in Eq.~\eqref{eq:H_0_def} are chosen to be zero in the middle of the 
single particle energy band, $\xi=0$.
\bibitem{footnote} Part of the interaction is decomposed in the 
superfluid channel, while the other part in the ferromagnetic
channel, and thus part of the interaction energy is apparently 
dropped.
\bibitem{footnote2} Of course, particle-hole symmetry can be broken in
 many other ways, too (by introducing an asymmetrical cut-off,
 $W_\pm$, e.g.), but the non-vanishing slope of the DOS seems to have the
 largest impact.
\bibitem{footnoteOrderParametersConvDifference} We remark that our definition of the 
    SF order parameter $\Delta_{\alpha\beta}$ differs from the 
    usual convention by a factor of 1/2. This comes from the factor 1/2 
    difference in our convention for the interaction parameter $\lambda_{\alpha\beta}$
    in Eq.~\eqref{eq:H_0_def}.
\bibitem{footnote_normal_order}
We verified that, indeed, this normal ordering provides the correct 
densities at the free energy minima.
\bibitem{footnoteTcStar} In case of a linear DOS, $T_c^{(*)}$ is slightly greater than $T_c$ for positive 
$\mu_y$'s, while it is slightly smaller than $T_c$ for $\mu_y<0$.
\bibitem{footnote3} Since the simultaneous shift of
  all the chemical potential components shall have no impact apart from
  changing the value of the Ginzburg-Landau coefficients, 
  we restrict ourselves to $\mathrm{Tr} \delta\tilde{\mathbf{\mu}}=0$.
\bibitem{footnote4} Notice that the term 
  $(\mathrm{Tr} (\widetilde{\mathbf{\Delta}}\widetilde{\mathbf{\Delta}}^+))^2$
  is proportional to 
  $\mathrm{Tr}((\widetilde{\mathbf{\Delta}}\widetilde{\mathbf{\Delta}}^+)^2)$, 
  and does not appear in the expansion.
\bibitem{footnote6} In the absence of particle-hole symmetry, one may
  need to rescale the fields $\Delta_{23}$ and $\Delta_{13}$ to have
  the same coefficient in the kinetic part.
\bibitem{footnote7} \marci{We note, that Eq.~\eqref{eq:WardGaussian} holds only at the 
solutions of the EOM self-consistency equations.}
\bibitem{footnote8} \marci{ We assume that
for each counter-propagating momenta the single particle energies
are the same, $\epsilon_{\mathbf{k}}=\epsilon_{-\mathbf{k}}$. (This
follows naturally from the inversion symmetry of the system.) 
This property makes the calculations in the Cooper channel
simple, since in each vertex 
both of the energies of the incoming lines, and both of 
those of the outgoing lines are either in $\mathcal{B}^{exp}$, or
neither of them is. Therefore the energies
that we integrate out can be trivially separated
in the calculation of ladder diagrams. }
\end{thebibliography}
\end{document}